\documentclass[
	aps,
	prd,
	reprint,
	groupedaddress,
	showpacs,
	nofootinbib,
	nobibnotes,
	amsmath,
	amssymb,
	]{revtex4-1}

\usepackage{epsfig}
\usepackage{graphicx}
\usepackage{url}
\usepackage{siunitx}
\usepackage[mathscr]{euscript}
\usepackage[linktocpage]{hyperref}

\usepackage{xcolor}
\usepackage{bm}
\def\vec{\bm}
\def\mat{\bm}

\begin{document}
\graphicspath{{Figs/}}
\title{Using inpainting to construct accurate cut-sky CMB estimators}
\author{H.F.~Gruetjen$^{1}$}
\email{helge.gruetjen@gmail.com}
\author{J.R.~Fergusson$^{1}$}
\author{M.~Liguori$^{2}$}
\author{E.P.S.~Shellard$^{1}$}

\affiliation{$^{1}$Centre for Theoretical Cosmology, DAMTP, University of Cambridge, Cambridge CB3 0WA, United Kingdom}
\affiliation{$^{2}$Dipartimento di Fisica e Astronomia G. Galilei,Universit\`a degli Studi di Padova, via Marzolo 8, I-35131 Padova, Italy}

\date{\today}

\begin{abstract}
The direct evaluation of manifestly optimal, cut-sky CMB power spectrum and bispectrum estimators is numerically very costly, due to the presence of inverse-covariance filtering operations. This justifies the investigation of alternative approaches. In this work, we mostly focus on an inpainting algorithm that was introduced in recent CMB analyses to cure cut-sky suboptimalities of bispectrum estimators. First, we show that inpainting can equally be applied to the problem of unbiased estimation of power spectra. We then compare the performance of a novel inpainted CMB temperature power spectrum estimator to the popular apodised pseudo-$C_l$ (PCL) method and demonstrate, both numerically and with analytic arguments, that inpainted power spectrum estimates significantly outperform PCL estimates. Finally, we study the case of cut-sky bispectrum estimators, comparing the performance of three different approaches: inpainting, apodisation and a novel low-l leaning scheme. Providing an analytic argument why the local shape is typically most affected we mainly focus on local type non-Gaussianity. Our results show that inpainting allows to achieve optimality also for bispectrum estimation, but interestingly also demonstrate that appropriate apodisation, in conjunction with low-l cleaning, can lead to comparable accuracy.
\end{abstract}


\maketitle
\tableofcontents

\section{Introduction}
\label{sec:Intro}
The study of the cosmic microwave background radiation (CMB) has become one of the most important areas in cosmology and is largely responsible for moving the field into a precision era. As the CMB anisotropies are almost Gaussian, the overwhelming majority of the information on cosmology is contained in the two-point correlation function, with significant constraints on the magnitude of deviations from Gaussianity coming from higher-order correlators like the bispectrum and trispectrum.

A central issue when analysing the CMB is how to deal with foreground contamination of the data. There are many sophisticated techniques cleaning the maps (see e.g.\ Ref.~\cite{Ade:2013CompSep,Adam:2015CompSep}) but in strongly contaminated regions, the only solution is to mask that part of the sky. While this approach is very effective at removing the contamination it is not without drawbacks. Masking the sky couples the multipoles together and allows power to leak between them. Exact inverse-covariance weighting accounts for the resulting correlations and guarantees minimal error bars of estimates \cite{TegmarkBunn:bruteforce, Gorski:COBEanalysis, Borrill:MADCAP, Bond:estCMB, Tegmark:losslessmeasure, Efstathiou:MLanalysis}. However, it is computationally extremely challenging for large state-of-the-art datasets and typically approximate methods like the pseudo-$C_l$ approach for power spectrum analysis \cite{WandeltHivonGorski:PCLmethod, Hivon:MASTER, Efstathiou:MythsandTruths, Hinshaw:FirstYearWMAP, Ade:2013Likelihood} are chosen in practice%
\footnote{Promising alternative approaches have been developed that might lead to feasible routes to exact inverse-covariance weighting \cite{Smith:GravLensInvVarWeight, Elsner:EffWiener,Elsner:FastFishCalc, Elsner:LikeFishSys}. We will not discuss these further here but rather focus on less costly alternative approaches.}. %
Similar approximations to optimal estimators are also used for bispectrum analysis. In both the power spectrum and the bispectrum case these methods typically suffer from leakage induced suboptimalities. There are two standard approaches that can ameliorate this effect. Apodisation (smoothing) of the mask is currently the method of choice for power spectrum analysis while inpainting masked regions is used for the study of higher-order correlators like the bispectrum. Note that inpainting in the sense used in this work is not an attempt to reconstruct the CMB in contaminated regions, a procedure sometimes also referred to as inpainting (see e.g. Refs.~\cite{Perotto:CMBreconstruction, Starck:lowlinpainting,BenoitLevy:lensingreconstruction}). We employ a simple linear inpainting procedure with the sole purpose of smoothing the sharp edges of the mask by assigning suitable values to masked pixels. Inpainting in this sense was first introduced in the context of binned bispectrum estimation \cite{Bucher:2009binned,Bucher:inpainting} as an ad-hoc method that produced near-optimal constraints and subsequently used for the Planck 2013 and 2015 bispectrum analyses \cite{Ade:2013NonGaussianity, Ade:2015PrimNG}.

The main purpose of this paper is to develop an understanding of when and why inpainting is useful and to compare its performance to other approaches. We will do so by first focusing on the temperature power spectrum in Sec.~\ref{sec:InpvsApo}, which is easier to study. In the process, we show that it is possible to construct analytically debiased power spectrum estimates from inpainted maps that significantly outperform PCL estimators (both unapodised and apodised). We also present improved PCL covariance approximations and generalise them to the case of power spectrum estimates from inpainted maps. It will become clear that inpainting is an asymmetric method in the sense that low-$l$-high-$l$ (low-high) coupling is highly suppressed, while high-$l$-low-$l$ (high-low) coupling is in fact exacerbated compared to unapodised PCL estimation. This asymmetry makes inpainting ideally suited for the analysis of strongly decaying power spectra where low-high coupling is highly problematic, but high-low coupling is largely irrelevant.

We go on to study the case of the bispectrum in Sec.~\ref{sec:bispecinp}. We review standard approximations to optimal cut-sky bispectrum estimators and analytically estimate the degradation of the estimator variance due to masking of the data in the case of an analysis based on a simple binary mask. The estimate shows that the impact can be very significant depending on the shape of the bispectrum under consideration. It also provides an analytic argument explaining the fact that the local shape is most affected (see e.g.\ \cite{Fergusson:Bispectrum2010,FergussonShellard:OptEst}). Focusing on the local shape as the worst-case scenario, we compare the performance of inpainting and apodisation. We also introduce a cleaning scheme that explicitly subtracts the leakage from low-$l$ power into high-$l$ modes and can be thought of as an exaggerated form of inverse-covariance weighting. This cleaning scheme enables the identification of leakage from low-$l$ modes as the primary origin of estimator suboptimalities which is also the picture suggested by the analytic estimate.

\section{Inpainting vs. apodisation: the power spectrum case}
\label{sec:InpvsApo}

\subsection{Setup}
\label{subsec:setup}
To ensure that a comparison of approaches accurately reflects the performances expected in a realistic analysis setting, we adopt a setup that mimics the Planck 2013 SMICA map \cite{Ade:2013CompSep}. As mentioned above, inpainting was first employed to extract bispectrum constraints from this map in the context of the 2013 analysis \cite{Ade:2013NonGaussianity}. The fiducial power spectrum we employ is a lensed $\Lambda$CDM model based on the 2013 best-fit parameter values. We multiply by the beam window of a Gaussian beam with  $5\,\mathrm{arcmin}$ FWHM and add white isotropic noise at a level chosen to match the SMICA map. Furthermore, we employ the 2013 U73 mask. We emphasise that despite the fact that the choice of beam, noise level and mask are motivated by the SMICA analysis, the results presented here obviously apply more generally.

\subsection{Error properties of pseudo-\texorpdfstring{$C_l$}{Cell} estimates}
\label{subsec:PCLandApo}
Given a mask function $U(\hat{\vec{n}})$ the pseudo multipoles $\tilde{a}_{lm}$ are given in terms of the full-sky multipoles $a_{lm}$ by
\begin{equation}\label{eq:pseudomultipoles}
\tilde{a}_{l_1m_1}=\int\mathrm{d}^2n\,Y^*_{l_1m_1}(\hat{\vec{n}})\Delta T(\hat{\vec{n}})U(\hat{\vec{n}})=P_{l_1m_1l_2m_2}a_{l_2m_2}\,,
\end{equation}
where we defined
\begin{equation}
P_{l_1m_1l_2m_2}=\int\mathrm{d}^2n\,Y^*_{l_1m_1}(\hat{\vec{n}})Y_{l_2m_2}(\hat{\vec{n}})U(\hat{\vec{n}})\,.
\end{equation}
Rather than evaluating optimal estimators that involve inverse-variance weighting of the $\tilde{a}_{lm}$, PCL estimation relies on the introduction of quantities $\tilde{C}_l$ analogous to standard full sky $C_l$ estimators given by
\begin{equation}
\tilde{C}_l=\frac{1}{2l+1}\sum_m \vert\tilde{a}_{lm}\vert^2\,.
\end{equation}
Using the full-sky covariance matrix it is easy to show that their expectation values are related to the true power spectrum $C_l$ via
\begin{equation}\label{eq:PCLrelation}
\langle \tilde{C}_{l_1}\rangle=\frac{1}{2l_1+1}\Pi_{l_1l_2}C_{l_2}+\langle\tilde{N}_{l_1}\rangle\,,
\end{equation}
where $\langle \tilde{N}_l\rangle$ is the noise contribution and%
\footnote{If we want to calculate the matrix $\mat{\Pi}$ exactly up to a certain $l_{\mathrm{max}}$ using the analytic formula on the RHS, the sum over $l_3$ in this expression should extend to $2\,l_{\mathrm{max}}$. Beyond that, the Wigner-3j symbol vanishes.}
\begin{align}\label{eq:couplingmat}\nonumber
\Pi_{l_1l_2}&=\sum_{m_1m_2}\left\vert P_{l_1m_1l_2m_2}\right\vert^2\\
&=(2l_1+1)(2l_2+1)\sum_{l_3}\frac{2l_3+1}{4\pi}U_{l_3}\left(\begin{array}{ccc} l_1 & l_2 & l_3 \\0&0&0\end{array}\right)^2
\end{align}
is a symmetrised version of the standard PCL coupling matrix (often denoted by $M_{l_1l_2}$) and $U_l$ is the power spectrum of the mask which is defined in terms of the multipole coefficients $u_{lm}$ of the mask function $U(\hat{\vec{n}})$ as
\begin{equation}
U_l=\frac{1}{2l+1}\sum_m \vert u_{lm}\vert^2\,.
\end{equation}
We deliberately defined the PCL coupling matrix $\mat{\Pi}$ without including the factor $1/(2l_1+1)$ to highlight the symmetry of the PCL coupling. We can terminate the sum in Eq.~\eqref{eq:PCLrelation} at some sufficiently high $l_{\mathrm{max}}$ in which case $\mat{M}$ is a square matrix of dimension $l_{\mathrm{max}}+1$ and%
\footnote{We assume that $\mat{\Pi}$ is invertible, which is usually the case unless the sky coverage $f_{\mathrm{sky}}$ of the experiment is very small (cf.\ Ref~\cite{Efstathiou:MythsandTruths} and references within).} %
we can obtain unbiased PCL estimates $\hat{C}^{\mathrm{PCL}}_l$ as
\begin{equation}
\hat{C}^{\mathrm{PCL}}_{l_1}=(\Pi^{-1})_{l_1l_2}(2l_2+1)\left(\tilde{C}_{l_2}-\langle\tilde{N}_{l_2}\rangle\right)\,.
\end{equation}
The advantage of this approach is its computational simplicity. Little numerical effort is necessary to arrive at unbiased power spectrum estimates. The $\tilde{C}_l$ are trivial to calculate and the expression for the coupling matrix $\mat{\Pi}$, Eq.~\eqref{eq:couplingmat}, can be evaluated numerically in a very efficient manner. However, while it can give rise to accurate estimators it is in general not optimal. Masking of the sky couples different multipoles so that power from nearby $l$ also contributes to a given $\tilde{C}_l$. This effect is described by the coupling matrix $\mat{\Pi}$ and multiplication with its inverse corrects for the resulting bias. Nonetheless the leakage from nearby multipoles also affects the variance of estimates. While in the case of a constant power spectrum PCL estimation is equivalent to inverse-variance weighting \cite{Efstathiou:MythsandTruths, Gruetjen:TowardsEfficient}, for non-constant $C_l$ leakage of power between different $l$ leads to an increase in the variance and thus causes the estimator to be suboptimal. The coupling width of the mask can be reduced by smoothing the edges, a method referred to as apodisation. Apodisation reduces long-range leakage and hence can have a very positive impact on the variance of PCL estimates. However, requiring smooth edges of the mask comes at the price of a loss of sky fraction which generally leads to larger errors. It is clear that a balance must be struck between smoothing the mask to reduce mode coupling, and retaining as much sky fraction as possible. For the purpose of this paper we use a simple apodisation procedure inspired by the one used in the 2013 Planck analysis \cite{Ade:2013Likelihood}. Our method ensures that masked regions of the unapodised mask remain masked so that all masks use the same set of data allowing for a meaningful comparison. The procedure can be thought of as an approximate convolution of the mask with a Gaussian beam of a given FWHM and is described in detail in App.~\ref{app:ApoMethod}. The effect of this smoothing is strong suppression of mode coupling beyond a multipole separation $\Delta l\sim \pi/\mathrm{FWHM}$.

\begin{figure*}[!htb]
\centering
\includegraphics[width=.8\textwidth]{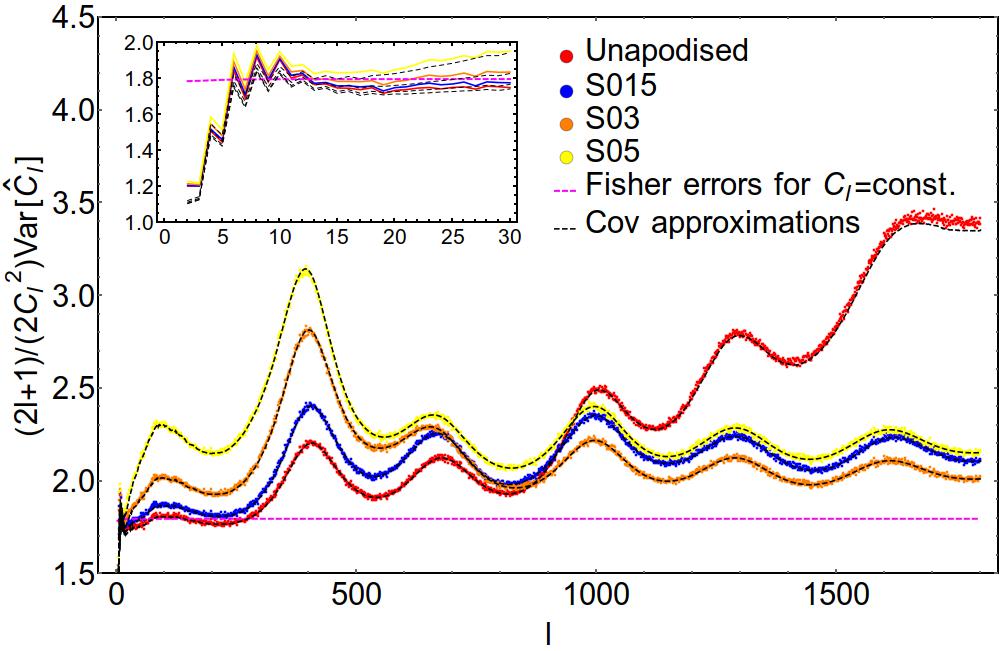}
\caption{The variances of PCL power spectrum estimates divided by full-sky variance for various degrees of apodisation. We plot exact variances as obtained from $10^5$ MC samples as well as variances predicted by analytic approximations to the PCL covariance matrices. The approximations perform very well with minor deviations only at the lowest $l$ and, in the case of the unapodised mask, highest $l$. Also plotted are Fisher errors obtained under the assumption of a constant power spectrum $C_l=C_0$ (flat magenta line). The inset magnifies the region at low $l$ where PCL variances approach the cosmic variance limit as expected based on the fact that the temperature CMB can be viewed as nearly band limited when studying the largest scales due to the large amount of power in low $l$ modes.}
\label{fig:PCLvar}
\end{figure*}

Figure~\ref{fig:PCLvar} shows a plot of PCL variances up to $l_{\mathrm{max}}=2000$ for the setup described in Sec.~\ref{subsec:setup}.
The plots are obtained using a HEALPix resolution parameter $N_{\mathrm{side}}=1024$. Besides the unapodised mask smoothing scales $0.15^{\circ}$, $0.3^{\circ}$ and $0.5^{\circ}$, labelled as S015, S03 and S05 respectively, are shown. It is evident how at low $l\lesssim 800$, long-range leakage due to the discontinuities is not as severe and unapodised PCL produces the lowest error bars. However, without any apodisation the error bars grow rapidly going to higher $l$. The S015 mask already eliminates most of the leakage but further improvements can be made using the S03 mask. Going to even more aggressive smoothing does not result in further improvements as the reduced leakage is not sufficient to counteract the loss of sky fraction that causes the error bars to grow everywhere. 

For comparison, Fig.~\ref{fig:PCLvar} includes the Fisher errors for the case of a constant power spectrum $C_l=\mathrm{const.}=C_0$. In this case unapodised PCL estimation is optimal and the Fisher errors can be easily evaluated as the covariance matrix is a simple projection operator and thus is its own pseudoinverse. They are simply given by
\begin{equation}
(F_{\mathrm{const.}}^{-1})_{ll}=2C_0^2(\Pi^{-1})_{ll}
\end{equation}
and as we roughly have%
$\Pi_{ll}\approx (2l+1)(f^u_{\mathrm{sky}})^2$ we get
\begin{equation}
\frac{2l+1}{2C_0^2}(F_{\mathrm{const.}}^{-1})_{ll}\approx (f^u_{\mathrm{sky}})^{-2}\approx1.85
\end{equation}
as a good estimate of the actual value plotted in Fig.~\ref{fig:PCLvar}. Because of leakage and reduced sky fraction in the apodised cases, the plotted ratios of PCL variance to full-sky variance are typically larger. A notable exception is the region at low $l$ that is also shown in the inset. The fact that the curves are well below the Fisher errors for $C_l=\mathrm{const.}$ is not a violation of the Cramer-Rao bound because we are comparing different fiducial spectra. As the $\Lambda$CDM temperature power spectrum decays rapidly at low $l$, the low-$l$ modes experience almost no leakage from the high-$l$ modes and thus can be estimated more accurately. Another way of saying this is that, because of the rapid drop in power, for the treatment of the modes at very low $l$, the power spectrum is approximately band-limited meaning that full-sky $a_{lm}$ can be reconstructed accurately from the cut-sky $\tilde{a}_{lm}$. Thus, the variance of the $C_l$ estimates can be close to full-sky cosmic variance and far below the $C_l=\mathrm{const.}$ Fisher errors (see e.g.\ Refs.~\cite{Efstathiou:MLanalysis, Pontzen:Cutskynotanomalous, Aurich:reconstructmaskedsky, Molinari:CompEstlargescales}).

\subsection{Pseudo-\texorpdfstring{$C_l$}{Cell} covariance approximations}
\label{subsec:PCLcovapp}
While we obtained variances from a large number of MC samples in the previous section, the exact covariance can also be obtained from the expression
\begin{align}\nonumber
&\mathrm{Cov}[\hat{C}^{\mathrm{PCL}}_{l_1},\hat{C}^{\mathrm{PCL}}_{l_2}]\\
=&2(\Pi^{-1})_{l_1l_3}(\Pi^{-1})_{l_2l_4}\sum\limits_{m_3m_4}\left\vert P_{l_3m_3l_5m_5}C_{l_5}P_{l_5m_5l_4m_4}\right\vert^2\,.
\end{align}
Evaluating this expression exactly is cumbersome but it is required to make the $C_l$-estimates useful for further analysis, in particular for the construction of a likelihood \cite{Ade:2013Likelihood,Efstathiou:MythsandTruths,Hamimeche:CMBlikelihoods}. In order not to spoil the computational simplicity of PCL estimation analytic approximations are thus used in practice. These approximations assume that the power spectrum is nearly constant so that the factors of $C_l$ can be removed from the sums \cite{Efstathiou:MythsandTruths, Challinor:ErrorAnaPol}. Following this approach consistently gives rise to an expression
\begin{equation}\label{eq:covappconst}
\mathrm{Cov}[\hat{C}^{\mathrm{PCL}}_{l_1},\hat{C}^{\mathrm{PCL}}_{l_2}]\approx 2C_{l_1}C_{l_2}(\Pi^{-1})_{l_1l_3}(\Pi^{-1})_{l_2l_4} \Pi^{(2)}_{l_3l_4}\,,
\end{equation}
where the superscript $(2)$ indicates that a quantity, in this case the coupling matrix, is evaluated for the square of the mask. Note that we made use of the completeness of the spherical harmonics to arrive at this expression. This approximation is inaccurate as it does not take the impact of variations in the power into account. For example in the case of an unapodised mask it simply reproduces the Fisher errors plotted in Fig.~\ref{fig:PCLvar} that evidently differ quite substantially from the exact result. To improve the approximation let us assume the general form to be the same as Eq.~\eqref{eq:covappconst}, but replace $C_l$ with an appropriate $\bar{C}_l$ that ensures that the approximation matches the true covariance well on the diagonal. More precisely we require
\begin{align}\label{eq:covapprox1}\nonumber
&\mathrm{Var}[\hat{C}^{\mathrm{PCL}}_{l}]\\\nonumber
=&2(\Pi^{-1})_{ll_1}(\Pi^{-1})_{ll_2}\sum\limits_{m_1m_2}\left\vert P_{l_1m_1l_3m_3}C_{l_3}P_{l_3m_3l_2m_2}\right\vert^2\\
\approx&2\bar{C}^2_{l}(\Pi^{-1})_{ll_1}(\Pi^{-1})_{ll_2} \Pi^{(2)}_{l_1l_2}\,.
\end{align}
Making use of the fact that all matrices involved are nearly diagonal we arrive at
\begin{equation}\label{eq:covapprox2}
\bar{C}^2_{l}\approx\frac{1}{\Pi^{(2)}_{ll}}\sum\limits_{m_1m_2}\left\vert P_{lm_1l_3m_3}C_{l_3}P_{l_3m_3lm_2}\right\vert^2\,.
\end{equation}
Now rather than simply treating the power spectrum as constant and pulling it out of the sums on the RHS let us try to find a better approximation. We can write
\begin{align}\label{eq:covapprox3}\nonumber
&\sum\limits_{m_1m_2}\left\vert P_{lm_1l_3m_3}C_{l_3}P_{l_3m_3lm_2}\right\vert^2\approx\sum\limits_{m}\left\vert P_{lml_3m}C_{l_3}P_{l_3mlm}\right\vert^2\\
\approx&\frac{A_l}{2l+1}\left(\sum\limits_{m}\vert P_{lml_3m}\vert^2C_{l_3}\right)^2=A_l(2l+1)\langle\tilde{C}_l\rangle^2\,,
\end{align}
where we treated the mask as azimuthally symmetric and introduced a factor $A_l$ that remains to be determined. The motivation for introducing $A_l$ in this way comes from the fact that if the terms in the sum are independent of $m$ we have $A_l=1$. Now we do not expect the terms to be completely independent of $m$ but we can assume that the effect of the variations can be well accounted for by evaluating $A_l$ in the case of a constant power spectrum, $C_l=C_0$, for which we have
\begin{align}\label{eq:covapprox4}\nonumber
&\sum\limits_{m}\left\vert P_{lml_3m}C_{l_3}P_{l_3mlm}\right\vert^2=C_0^2\sum\limits_{m}\left(P^{(2)}_{lmlm}\right)^2\\
=&C_0^2\Pi^{(2)}_{ll}=\frac{\Pi^{(2)}_{ll}}{(f^{(2)}_{\mathrm{sky}})^2}\langle\tilde{C}_0\rangle^2\,,
\end{align}
so that
\begin{equation}
A_l=\Pi^{(2)}_{ll}/((2l+1)(f^{(2)}_{\mathrm{sky}})^2)\,, 
\end{equation}
which is in fact typically independent of $l$ except at very small $l$ and relatively close but slightly larger than unity%
\footnote{The fact that it must be larger than unity can be viewed as a direct consequence of the Cauchy-Schwarz inequality.}. %
Substituting this result back into Eqs.~\eqref{eq:covapprox3} and \eqref{eq:covapprox2} we obtain
\begin{equation}\label{eq:covapprox5}
\bar{C}_{l}=\frac{\langle\tilde{C}_l\rangle}{f^{(2)}_{\mathrm{sky}}}\,.
\end{equation}
Summing up, the covariance approximation used in this work reads
\begin{equation}\label{eq:covapprox6}
\mathrm{Cov}[\hat{C}^{\mathrm{PCL}}_{l_1},\hat{C}^{\mathrm{PCL}}_{l_2}]\approx2\bar{C}_{l_1}\bar{C}_{l_2}(\Pi^{-1})_{l_1l_3}(\Pi^{-1})_{l_2l_4} \Pi^{(2)}_{l_3l_4}
\end{equation}
with $\bar{C}_l$ as in Eq.~\eqref{eq:covapprox5}.

These covariance approximations are also plotted in Fig.~\ref{fig:PCLvar} and the agreement with the exact variances as obtained from MC is very good (also see Fig.~\ref{fig:inpcovappcomp} below). A nice property of writing the covariances in this way is that the loss of sky fraction that increases errors everywhere mostly enters through the coupling matrices and leaves $\bar{C}_l$ largely unchanged. Effects on the variance due to variations in the power spectrum and the resulting leakage enter through deviations of $\bar{C}_l$ from $C_l$ with the two being equal in the case of a constant power spectrum (cf.\ Fig.~\ref{fig:barCl} and its discussion).

Note that leakage also affects the off-diagonal entries. Analytic approximations like the one above cannot fully capture the impact of variations of the power spectrum on correlations and generally deviate from the exact covariance matrix. However, elements far off the diagonal are typically very small so that errors are likely not important in practice (cf.~Ref.~\cite{Challinor:ErrorAnaPol}). 

\subsection{Inpainting as an alternative approach}
\label{subsec:InpPS}
\begin{figure*}[tb]
\centering
\includegraphics[width=.8\textwidth]{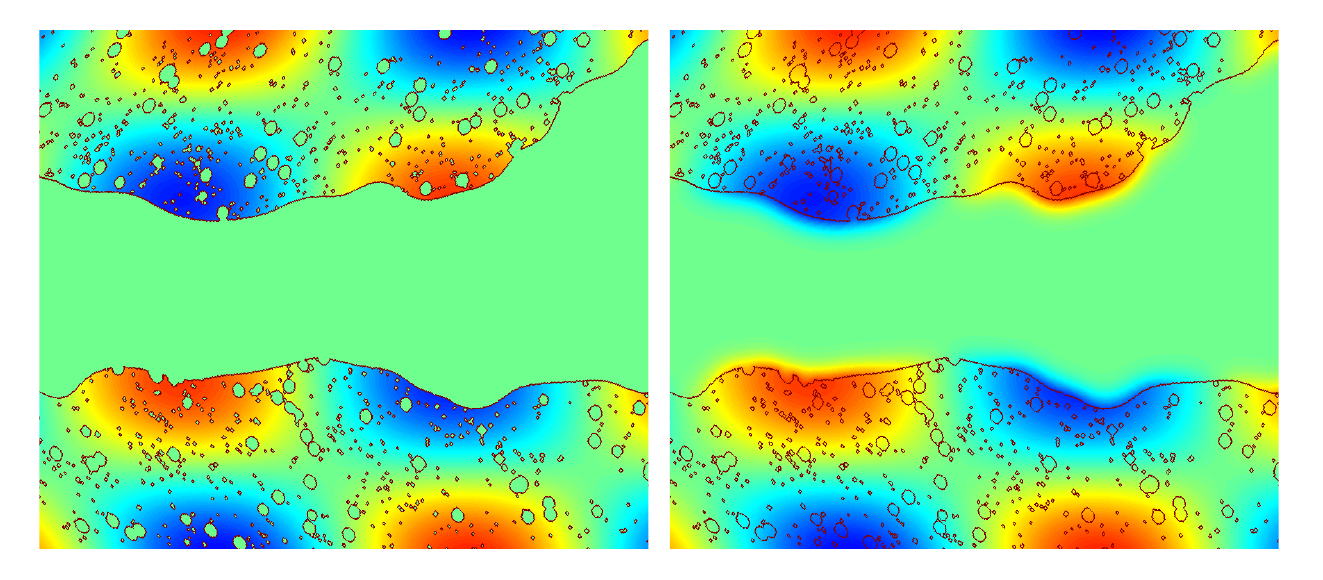}
\caption{Real part of the masked spherical harmonic $\mathrm{Re}\lbrace Y_{10\,5}\rbrace$ (left) and the corresponding inpainted spherical harmonic $\mathrm{Re}\lbrace Y^{\mathrm{I}}_{10\,5}\rbrace$. In both cases the boundary of the mask is highlighted.}
\label{fig:inpYlm}
\end{figure*}
Rather than apodising the mask function, inpainting multiplies the map with the unapodised mask and then fills in the excluded regions. For our purposes, the crucial requirements are that the method is linear and thoroughly eliminates any discontinuities introduced through masking of the data. As already mentioned in the introduction, {\em{inpainting in the sense used in this chapter does not refer to a procedure that attempts to reconstruct the masked regions of the CMB as in other work}}. The approach adopted here is a very simple routine identical to the one used in the Planck analysis \cite{Ade:2013NonGaussianity, Ade:2015PrimNG}.

Starting with a map of temperature fluctuations, $\Delta T(\hat{\vec{n}})$, and an unapodised mask function $U(\hat{\vec{n}})$, we first determine the set of unmasked pixels, for which $U(\hat{\vec{n}})=1$, and the set of masked pixels, for which $U(\hat{\vec{n}})=0$. The value of $\Delta T(\hat{\vec{n}})$ at unmasked pixels is left unchanged by the procedure and the inpainted map is still given by $\Delta T^{\mathrm{I}}(\hat{\vec{n}})=\Delta T(\hat{\vec{n}})$ for this set of pixels. The masked pixels are zeroed at the start of the procedure. Then we perform iterations on the set of masked pixels, where at each step a pixel gets assigned the average value of its immediate neighbours. The precise number of iterations is not directly relevant as long as it is large enough to ensure that the resulting map $\Delta T^{\mathrm{I}}(\hat{\vec{n}})$ is sufficiently smooth. Figure~\ref{fig:inpYlm} shows an example of the effect of inpainting. On the left we plot $\mathrm{Re}\lbrace Y_{10\,5}(\hat{\vec{n}})\rbrace$ for the masked spherical harmonic in a region that includes part of the galactic cut and various point source holes at HEALPix resolution $N_{\mathrm{side}}=512$. The right of the figure shows the real part of the inpainted spherical harmonic $\mathrm{Re}\lbrace Y^{\mathrm{I}}_{10\,5}(\hat{\vec{n}})\rbrace$ after $250$ inpainting iterations. In both plots the boundary of the mask is highlighted. We see that after inpainting the point source holes are filled in smoothly and the inpainted spherical harmonic now extends noticeably into the galactic cut. 

While this procedure seems largely ad-hoc, it meets both requirements stated above. In particular it is a manifestly linear operation. The inpainted multipoles $\tilde{a}^{\mathrm{I}}_{lm}$ are related to the full-sky multipoles $a_{lm}$ via
\begin{equation}
\tilde{a}^I_{l_1m_1}=I_{l_1m_1l_2m_2}a_{l_2m_2}\,,
\end{equation}
where we introduced the inpainting matrix $\mat{I}$. Its entries are simply the harmonic transforms of inpainted spherical harmonics $Y^{\mathrm{I}}_{l_2m_2}(\hat{\vec{n}})$, i.e.\
\begin{equation}\label{eq:inpaintingmatrix}
I_{l_1m_1l_2m_2}=\int\mathrm{d}^2n\,Y^*_{l_1m_1}(\hat{\vec{n}})Y^{\mathrm{I}}_{l_2m_2}(\hat{\vec{n}})\,.
\end{equation}

\subsection{Inpainting coupling matrix and power spectrum estimates}
\begin{figure*}[tb]
\centering
\includegraphics[width=.49\textwidth]{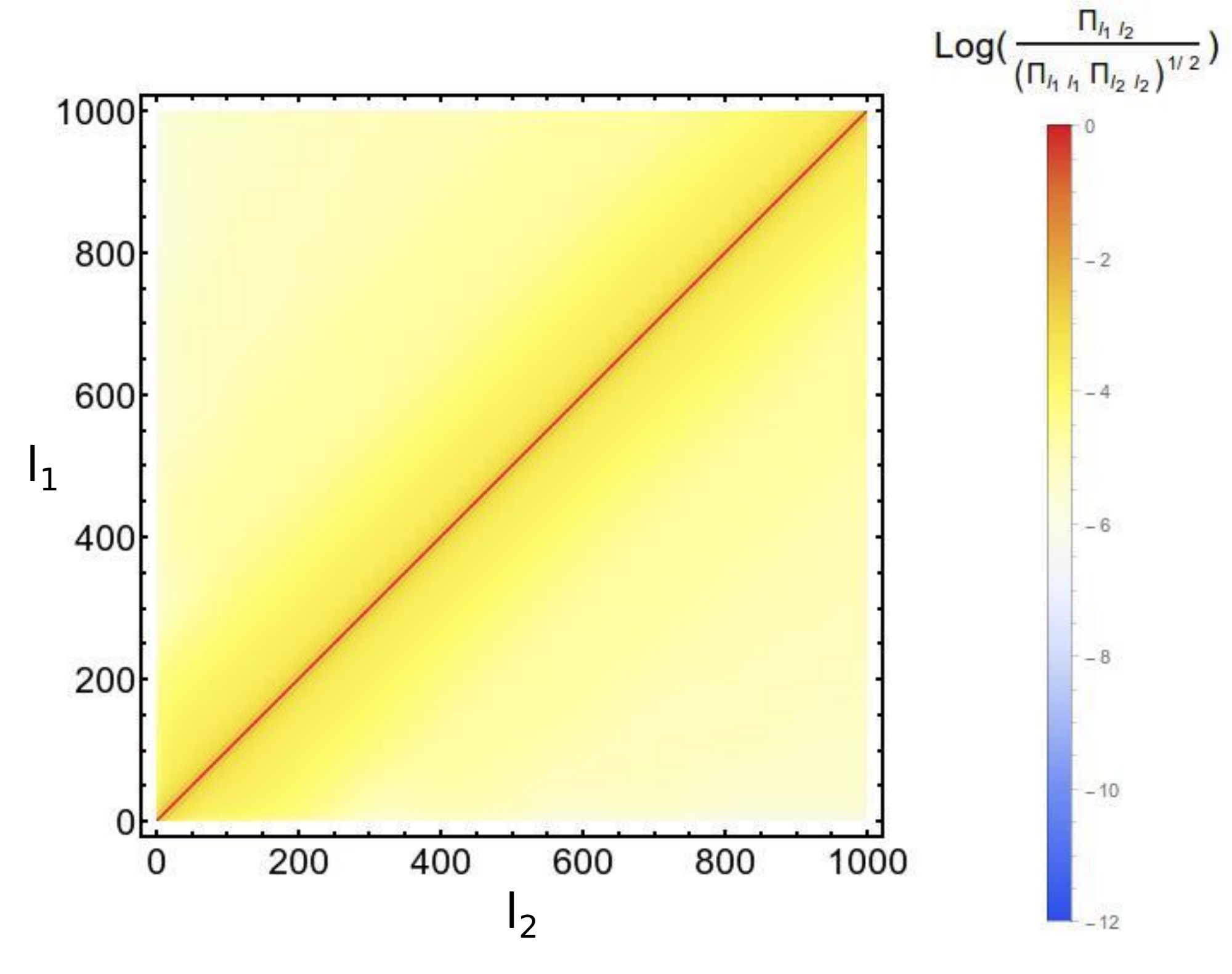}
\includegraphics[width=.49\textwidth]{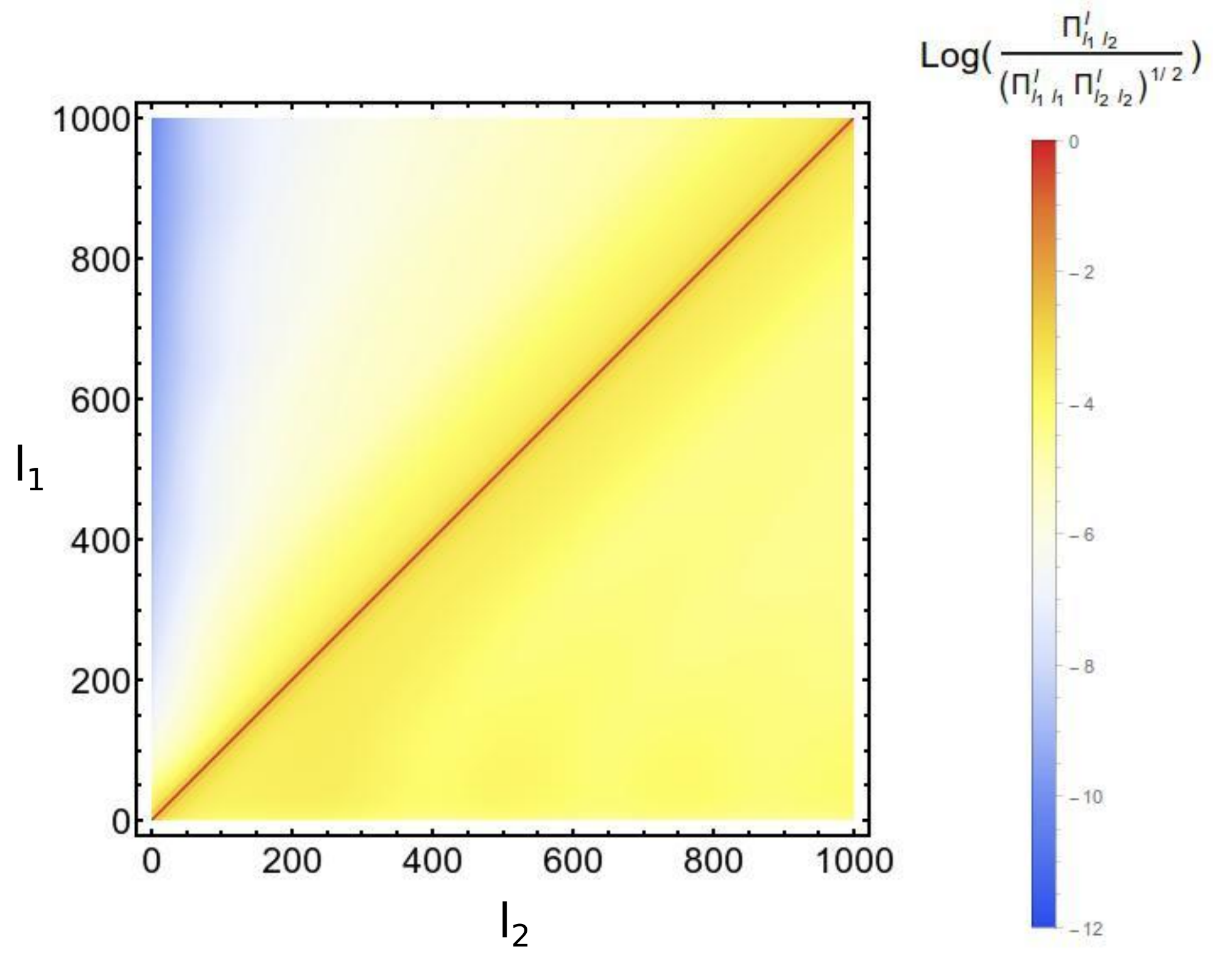}
\includegraphics[width=.7\textwidth]{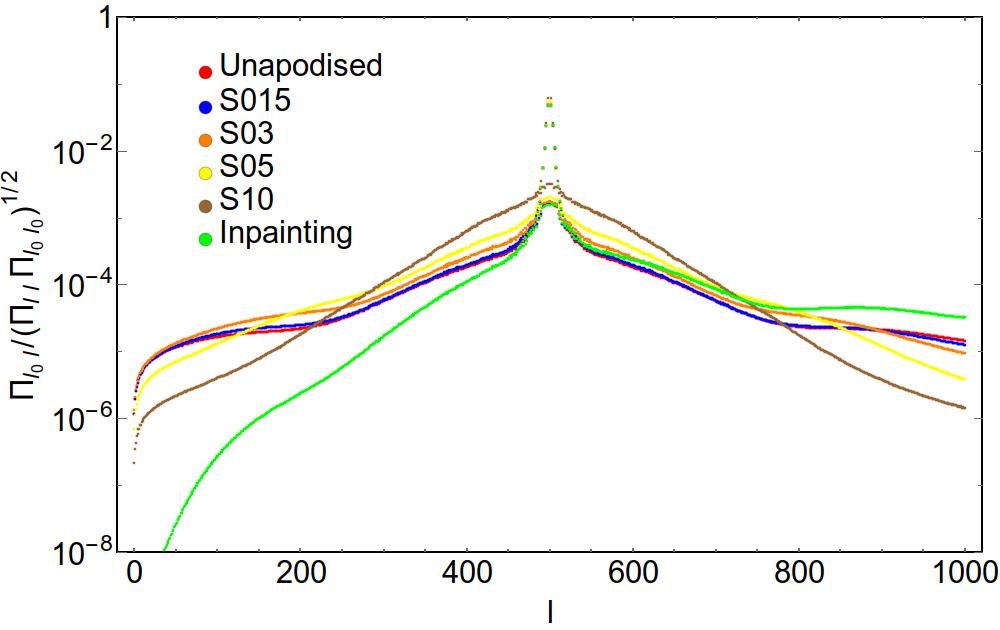}
\caption{Density plots of coupling matrices in the case of the unapodised U73 mask (top left) and inpainting (top right). The bottom panel shows a cross section $\Pi_{l_0l}$ through coupling matrices for various smoothing scales and inpainting at $l_0=500$. The asymmetry of the inpainting coupling is clearly visible. While low-high coupling is heavily suppressed compared to all PCL cases, high-low coupling is in fact enhanced as explained in the main text.}
\label{fig:couplingmats}
\end{figure*}
Writing the impact of inpainting as in Eq.~\eqref{eq:inpaintingmatrix} emphasises the similarities to standard PCL estimation. Comparing this equation to Eq.~\eqref{eq:pseudomultipoles}, we see that the inpainting matrix $\mat{I}$ simply replaces the spherical harmonic coupling matrix $\mat{P}$. Thus, in a similar fashion we can construct unbiased power spectrum estimates $\hat{C}^{\mathrm{I}}_l$ from inpainted maps via
\begin{equation}
\hat{C}^{\mathrm{I}}_{l_1}=\left((\Pi^{\mathrm{I}})^{-1}\right)_{l_1l_2}(2l_2+1)\left(\tilde{C}^{\mathrm{I}}_{l_2}-\langle\tilde{N}^{\mathrm{I}}_{l_2}\rangle\right)\,,
\end{equation}
where 
\begin{equation}
\tilde{C}^{\mathrm{I}}_l=\frac{1}{2l+1}\sum_m \vert\tilde{a}^{\mathrm{I}}_{lm}\vert^2
\end{equation}
and the inpainting coupling matrix $\mat{\Pi}^{\mathrm{I}}$ is given by
\begin{equation}
\Pi^{\mathrm{I}}_{l_1l_2}=\sum_{m_1m_2}\left\vert I_{l_1m_1l_2m_2}\right\vert^2\,.
\end{equation}
While this coupling matrix cannot be brought into a simpler form through analytic manipulations as in the PCL case, the steps involved in the calculation of $\mat{\Pi}^{\mathrm{I}}$ are trivial to parallelise and its numerical calculation is feasible on state-of-the-art supercomputers. The inpainting procedure has to be carried out for each spherical harmonic separately. For the purpose of this section we work with a HEALPix resolution parameter $N_{\mathrm{side}}=512$ and inpaint using $250$ iterations%
\footnote{The Planck analysis \cite{Ade:2013NonGaussianity, Ade:2015PrimNG} used $1000$ iterations at resolution $N_{\mathrm{side}}=2048$, which should be comparable as the smoothing scale due to inpainting scales as number of iterations times $N_{\mathrm{side}}$.}. %
In this case, calculation of the coupling matrix using the brute force inpainting algorithm outlined above and HEALPix for spherical transforms took $\mathcal{O}(10^3)$ CPU hours.

Figure~\ref{fig:couplingmats} visualises the coupling matrices of PCL estimation and inpainting. The top panels plot the scaled coupling matrices for the unapodised U73 mask (left) and the inpainting coupling matrix (right). The bottom panel shows a cross section through the coupling matrices $\Pi_{l_0l}$ at $l_0=500$ for various smoothing scales and inpainting. The plot can be interpreted as depicting the contribution of full-sky power at scale $l$ to the cut-sky power at $l_0$. The figure highlights the asymmetry of the inpainting coupling. It clearly shows how inpainting in fact increases the high-low coupling compared to the unapodised case, but is extremely effective at reducing low-high coupling.

The asymmetry shown in Fig.~\ref{fig:couplingmats} is the main advantage of inpainting. It is extremely efficient at eliminating leakage from modes at a given $l$ into modes at higher $l$. This can be intuitively understood. A full-sky spherical harmonic $Y_{lm}$ that is masked has power with much larger $l$ due to sharp features in the mask. However, inpainting renders any discontinuities very smooth so that the inpainted spherical harmonic has very little power at larger $l$ which suppresses low-high leakage. Conversely, inpainting is ineffective at reducing leakage from higher to lower $l$. A highly oscillatory masked spherical harmonic will acquire more low-$l$ power due to the inpainting procedure creating smoother contributions in masked regions. Hence, the high-low coupling is not reduced but rather exacerbated somewhat compared to the case of a simple unapodised mask. 

This asymmetry of the coupling makes inpainting ideally suited for CMB temperature analysis. Temperature power spectra decay rapidly so high-low leakage is generally irrelevant while low-high leakage is very important. As data in unmasked regions is unaffected by the inpainting procedure, the reduction of leakage does not come at the cost of a loss of sky fraction.

\subsection{Variances of inpainted power spectrum estimates}
\begin{figure*}[tb]
\centering
\includegraphics[width=.8\textwidth]{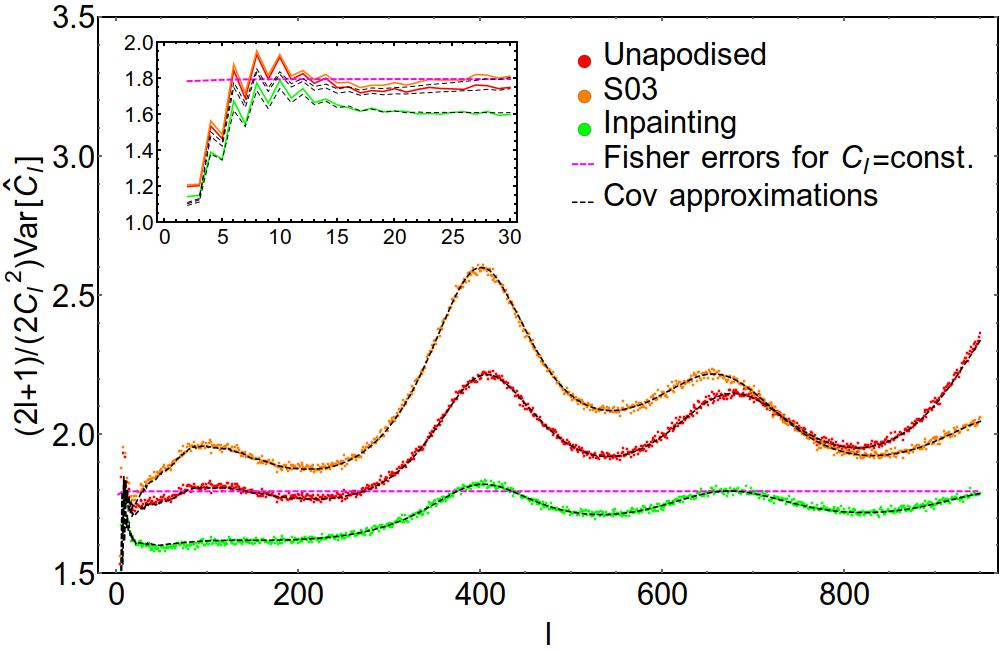}
\caption{The variances of inpainted power spectrum estimates divided by full-sky variance. For comparison the PCL variances for the unapodised and S03 mask and the Fisher errors obtained under the assumption of a constant power spectrum are included as well. As in Fig.~\ref{fig:PCLvar} we plot exact variances as obtained from $10^5$ MC samples as well as variances predicted by analytic approximations to the covariance matrices. Just like the PCL approximations, the analytic approximation to the inpainting variance discussed in Sec.~\ref{subsec:appinpcov} performs very well. The inset magnifies the low-$l$ region. Note that upon close inspection very minor differences at high $l$ between the PCL variances shown here and in Fig.~\ref{fig:PCLvar} are visible that arise because of the different choices of $N_{\mathrm{side}}$ that cause resolution effects in the harmonic transforms and the apodisation scheme.}
\label{fig:inpvar}
\end{figure*}

We saw in Sec.~\ref{subsec:PCLandApo} that, even after apodisation, low-high leakage leads to increases in the variance of PCL estimates whenever the power spectrum exhibits a drop. With the insight that inpainting is very efficient at suppressing this type of leakage, we expect that improvements can be made. The variance of inpainted $C_l$-estimates along with PCL variances in the unapodised and $0.3^{\circ}$-apodised case are shown in Fig.~\ref{fig:inpvar}.
The figure shows variances obtained from $10^5$ MC samples in each case. Also plotted are Fisher errors in the case of a constant power spectrum, $C_l=\mathrm{const.}=C_0$, already shown in Fig.~\ref{fig:PCLvar}. Inpainting significantly outperforms any of the apodised PCL estimates. We emphasise again that a variance below the Fisher errors for $C_l=\mathrm{const.}$ is not a violation of the Cramer-Rao bound because of the different underlying fiducial models. Rather, this phenomenon can be understood intuitively as reflecting the idea that when the power spectrum has a drop beyond a given $l$, it is possible to measure the power better than in the case of a flat spectrum because the contribution due to leakage from higher-$l$ modes is reduced. For example, the dips at $l\sim 500$ and $l\sim 800$ correspond to the locations of the acoustic peaks where the second derivative of the power spectrum briefly becomes negative.

\begin{figure}[tb]
\centering
\includegraphics[width=\columnwidth]{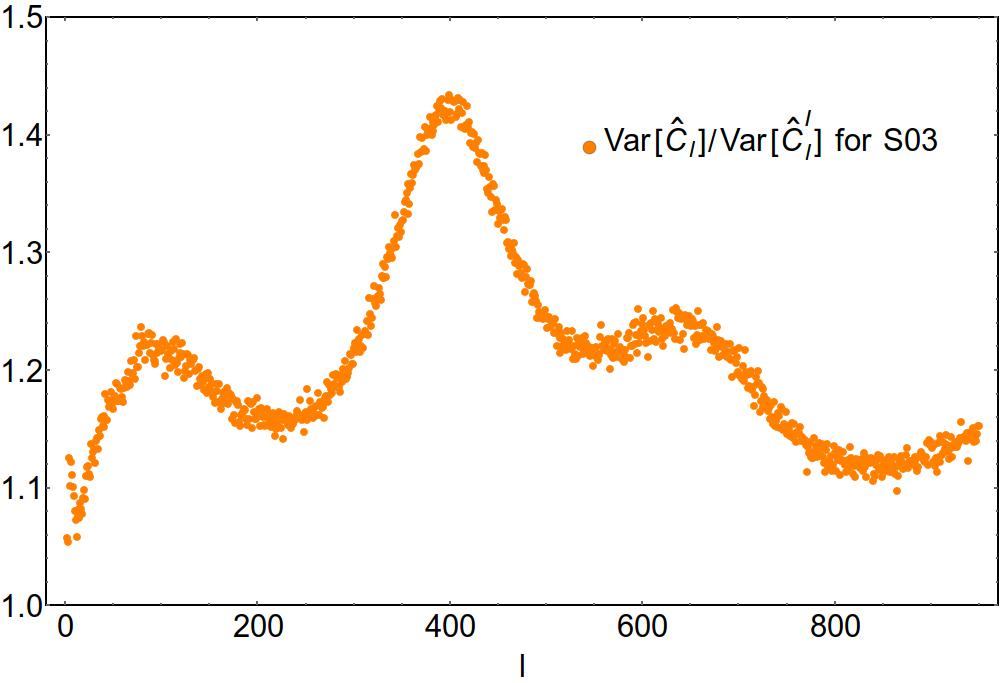}
\caption{Plot of the ratio of the variance of PCL estimates in the case of the S03 mask and the inpainted power spectrum estimates.}
\label{fig:inpvarratio}
\end{figure}

Figure~\ref{fig:inpvarratio} shows the ratio of the variance of the PCL estimates in the case of the S03 mask and inpainted power spectrum estimates to highlight the improvements. 
Inpainting leads to gains exceeding $15\%$ nearly across the full range with a maximum improvement of about $45\%$ seen at $l=400$ after the first acoustic peak of the power spectrum.

\subsection{Approximations to the inpainting covariance matrix}
\label{subsec:appinpcov}
\begin{figure}[t]
\centering
\includegraphics[width=\columnwidth]{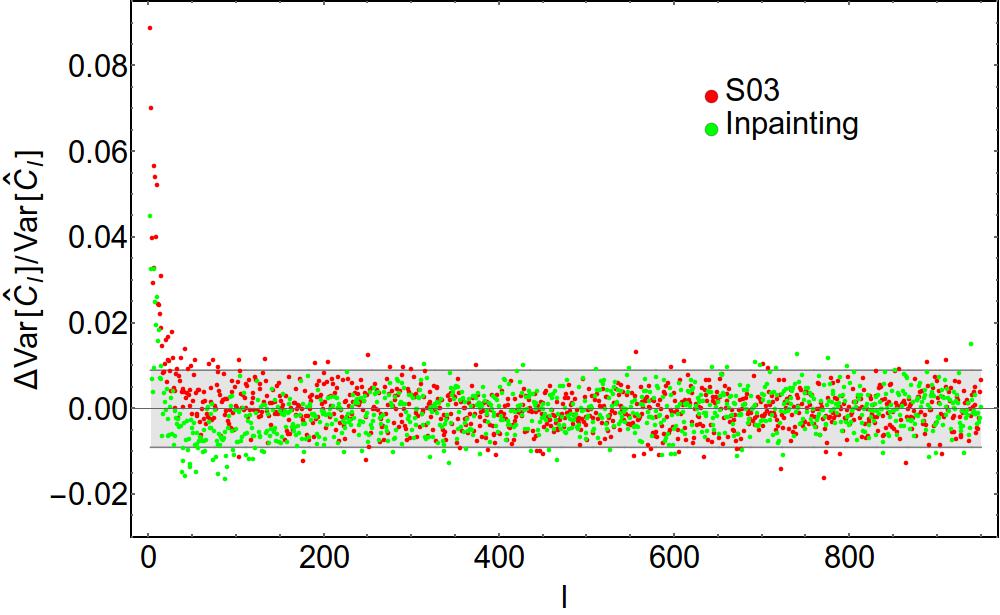}
\caption{The relative difference between the covariance approximations and variances obtained from MC sampling the case of PCL estimation using the S03 mask and inpainted power spectrum estimates. The shaded area indicates $2\sigma$ error bars on the variances expected from $10^5$ MC samples. The only notable differences occur at very low $l$.}
\label{fig:inpcovappcomp}
\end{figure}

Evaluation of the covariance matrix of inpainted power spectrum estimates, either through direct calculation or through extensive MC sampling, is computationally very expensive. As accurate covariance matrices are needed for the construction of a likelihood, it is desirable to obtain an analytic approximation. Building on the discussion in Sec.~\ref{subsec:PCLcovapp}, we attempt to find a suitable generalisation of the improved approximations to the covariance matrices of PCL estimates to the case of inpainting. Let us assume that we can approximate the covariance in a similar fashion as the unapodised PCL case, i.e.\
\begin{equation}
\mathrm{Cov}[\hat{C}^{\mathrm{I}}_{l_1},\hat{C}^{\mathrm{I}}_{l_2}]\approx2\bar{C}^{\mathrm{I}}_{l_1}\bar{C}^{\mathrm{I}}_{l_2}(\Pi^{-1})_{l_1l_2}\,,
\end{equation}
where $\mat{\Pi}$ is the coupling matrix of unapodised PCL estimates. We then only need to determine the correct choice of $\bar{C}^{\mathrm{I}}_l$. For the variance of the inpainted estimates we have
\begin{align}\label{eq:appcovapprox7}\nonumber
&\mathrm{Var}[\hat{C}^{\mathrm{I}}_{l}]\\\nonumber
=&2((\Pi^{\mathrm{I}})^{-1})_{ll_1}((\Pi^{\mathrm{I}})^{-1})_{ll_2}\sum\limits_{m_1m_2}\left\vert I_{l_1m_1l_3m_3}C_{l_3}I^*_{l_2m_2l_3m_3}\right\vert^2\\
\approx& 2(\bar{C}^{\mathrm{I}}_{l})^2(\Pi^{-1})_{ll}\,.
\end{align}
To evaluate this expression further, we make use of the fact that all matrices are nearly diagonal and obtain  
\begin{equation}\label{eq:appcovapprox8}
(\bar{C}^{\mathrm{I}}_{l})^2\approx\frac{\Pi_{ll}}{(\Pi^{\mathrm{I}}_{ll})^2}\sum\limits_{m_1m_2}\left\vert I_{lm_1l_3m_3}C_{l_3}I^*_{lm_2l_3m_3}\right\vert^2\,.
\end{equation}
Now, evaluating the sum on the RHS exactly in the same way as we did in Sec.~\ref{subsec:PCLcovapp} in the case of approximations to PCL covariance matrices,
\begin{equation}
\sum\limits_{m_1m_2}\left\vert I_{lm_1l_3m_3}C_{l_3}I^*_{lm_2l_3m_3}\right\vert^2=A_l(2l+1)\langle\tilde{C}^{\mathrm{I}}_l\rangle^2\,,
\end{equation}
and using the previous result for $A_l$ for an unapodised mask (recall that in this case $f^{(2)}_{\mathrm{sky}}\equiv f^{\mathrm{u}}_{\mathrm{sky}}$ and $\Pi^{(2)}_{ll}\equiv \Pi_{ll}$),
\begin{equation}
A_l=\frac{\Pi_{ll}}{(2l+1)(f^{\mathrm{u}}_{\mathrm{sky}})^2}\,,
\end{equation}
we arrive at
\begin{equation}\label{eq:appcovapprox9}
\bar{C}^I_l=\frac{\Pi_{ll}}{\Pi^{\mathrm{I}}_{ll}}\frac{\langle\tilde{C}^I_{l}\rangle}{f^{\mathrm{u}}_{\mathrm{sky}}}\,.
\end{equation}
The resulting analytic approximation to the variance of the inpainted $C_l$-estimates is plotted in Fig.~\ref{fig:inpvar} along with the corresponding approximations to the PCL variances already shown in Fig.~\ref{fig:PCLvar}. We see that the inpainting approximation agrees very well with the variances obtained from MC sampling. This is confirmed by Fig.~\ref{fig:inpcovappcomp} that shows the relative difference between the diagonals of the covariance approximations and the MC variances in the case of the S03 PCL estimates and the inpainted estimates.
Both approximations show agreement at the subpercent level with most of the difference being due to the scatter of the MC variances. The plot also includes $2\sigma$ error bars highlighting that the only significant deviations are observed at very low $l$. In practice, this region is not very relevant anyway as typically an exact low-$l$ likelihood replaces the fiducial Gaussian approximation there (e.g.\ in the Planck analysis \cite{Ade:2013Likelihood}). Obviously, the good agreement on the diagonal does not necessarily imply that the off-diagonal agreement is equally satisfactory. To make this approximation reliable and well suited for the construction of a likelihood, we need to assume that at least the near-diagonal correlations of $C_l$-estimates from inpainting can be modelled as being similar to the unapodised PCL case. To what extent this is true and whether the approximation is sufficiently accurate for the construction of a likelihood is beyond the scope of this chapter.

\begin{figure*}[t]
\centering
\includegraphics[width=.6\textwidth]{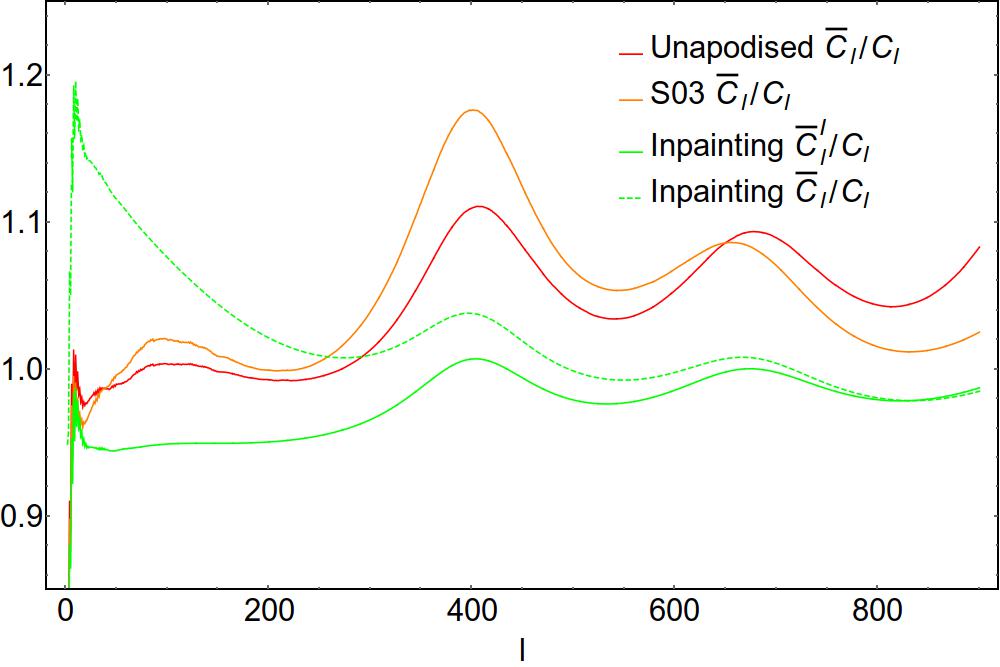}
\caption{The ratios $\bar{C}_l/C_l$ for PCL estimation using the unapodised and S03 mask along with $\bar{C}^{\mathrm{I}}_l/C_l$ for inpainting. The plot also shows $\bar{C}_l$ in the case of inpainting to highlight the impact of the $\Pi_{ll}/\Pi^{\mathrm{I}}_{ll}$ scaling entering $\bar{C}^{\mathrm{I}}_l$.}
\label{fig:barCl}
\end{figure*}

Before we conclude this section, we will briefly pause to elaborate further on the factors $\bar{C}_l$ entering the covariance matrices and in particular highlight the difference between $\bar{C}_l$ for inpainting and the related quantity $\bar{C}^{\mathrm{I}}_l$. Just as in the case of its PCL equivalents, the approximation to the covariance of inpainted power spectrum estimates is affected by the sky fraction mainly through the coupling matrix $\mat{\Pi}$ that scales as $(f^{\mathrm{u}}_{\mathrm{sky}})^{-2}$, while the shape of the power spectrum enters through deviations of $\bar{C}^{\mathrm{I}}_l$ from $C_l$. Figure~\ref{fig:barCl} plots $\bar{C}^{\mathrm{I}}_l$ for inpainting along with $\bar{C}_l$ for PCL estimation in the case of the unapodised and S03 mask.
Unsurprisingly, the curves closely resemble the variance plots shown in Fig.~\ref{fig:inpvar}. Let us first focus on the PCL cases. As the loss of sky fraction is divided out in the definition of the $\bar{C}_l$, the S03 curve shifts down with respect to the unapodised curve when comparing the two figures. Since the long-range coupling is significantly reduced, the S03 case has significantly lower $\bar{C}_l$ at high $l$, which was the main motivation to introduce apodisation in the first place. However, this plot shows that the increase in the S03 variance compared to the unapodised case in Fig.~\ref{fig:PCLvar} at smaller $l$ is not simply due to a loss of sky fraction, but due to an increase in mode coupling of short range. This is expected from Fig.~\ref{fig:couplingmats} which indeed shows an increase in short-range coupling. It can be intuitively understood as being due to the introduction of features in the mask with angular scale given by the smoothing scale which should increase the coupling below separations $\Delta l\sim \pi/\mathrm{FWHM}$.

Moving on to the inpainting curves, the difference between $\bar{C}_l$ and $\bar{C}^{\mathrm{I}}_l$ due to the factor $\Pi_{ll}/\Pi^{\mathrm{I}}_{ll}$ entering $\bar{C}^{\mathrm{I}}_l$ is particularly evident at lower $l$. $\bar{C}_l$ for inpainting is significantly higher than the corresponding PCL curves at low $l$, meaning that the variance of the inpainted pseudomultipoles $\tilde{a}^{\mathrm{I}}_{lm}$ is larger than their PCL equivalents $a_{lm}$ at low $l$. This effect is not due to leakage but rather can be thought of as the result of amplification of power at low $l$. Consider the simplest example of having only a monopole $a_{00}$ in the full-sky map. Unapodised PCL estimation would measure a pseudomultipole $\tilde{a}_{00}=f^{\mathrm{u}}_{\mathrm{sky}}a_{00}$ but the corresponding inpainted multipole $\tilde{a}^{\mathrm{I}}_{00}$ must be larger, $|\tilde{a}^{\mathrm{I}}_{00}|>|\tilde{a}_{00}|$, as inpainting amplifies the monopole by partially filling in the masked regions. Hence, there is an artificial amplification of power due to inpainting at low $l$ that is divided out in the final estimates. This effect is taken into account by the factor $\Pi_{ll}/\Pi^{\mathrm{I}}_{ll}$ and only $\bar{C}^{\mathrm{I}}_l$ enters the variance of the final inpainted estimates%
\footnote{Note that at higher $l$ inpainting may also have the opposite effect. Upon close inspection, it can be seen in Fig.~\ref{fig:barCl} that $\langle\tilde{C}^{\mathrm{I}}_l\rangle/f^{\mathrm{u}}_{\mathrm{sky}}$ starts to dip slightly below $\bar{C}^{\mathrm{I}}_l$ beyond $l\sim 800$, indicating that inpainted high-$l$ spherical harmonics $Y^{\mathrm{I}}_{lm}$ have reduced power at scale $l$.}.

\section{The case of the bispectrum}
\label{sec:bispecinp}
In Sec.~\ref{sec:InpvsApo} we focused on the power spectrum to investigate when and why inpainting is a useful approach for CMB analysis. The fact that it strongly reduces mode coupling makes it very attractive for the analysis of higher-order correlators such as the bispectrum as well. As mentioned above, inpainting was actually first introduced for bispectrum analysis by the Planck team in Ref.~\cite{Ade:2013NonGaussianity}. Building on the insights obtained by studying the power spectrum, we now proceed to study the case of the bispectrum. We will first attempt to obtain an analytic understanding of the impact of leakage on bispectrum estimates. Then, focusing on the local-shape bispectrum estimator that is particularly sensitive to low-high coupling, we go on to study and compare the ability of inpainting and alternative methods to restore optimality of cut-sky approximations to the optimal estimator.

\subsection{Optimal bispectrum estimator}
\label{subsec:optbispecest}
Suppose we want to estimate a small parameter $f_{\mathrm{NL}}$ that, at leading order, gives rise to deviations of the PDF $\mathcal{P}\left(\tilde{\vec{a}}\right)$ from Gaussianity of the form
\begin{align}\nonumber
&\langle \tilde{a}_{l_1m_1}\tilde{a}_{l_2m_2}\tilde{a}_{l_3m_3}\rangle=f_{\mathrm{NL}}\,\tilde{B}^{l_1l_2l_3}_{m_1m_2m_3}\\
\equiv&f_{\mathrm{NL}}\,P_{l_1m_1l_4m_4}P_{l_2m_2l_5m_5}P_{l_3m_3l_6m_6}B^{l_4l_5l_6}_{m_4m_5m_6}\,.
\end{align}
Here, $\tilde{B}^{l_1l_2l_3}_{m_1m_2m_3}$ is the cut-sky bispectrum and $B^{l_1l_2l_3}_{m_1m_2m_3}$ is the full-sky bispectrum. The latter is given in terms of the reduced bispectrum $b_{l_1l_2l_3}$ via $B^{l_1l_2l_3}_{m_1m_2m_3}=\mathcal{G}^{l_1l_2l_3}_{m_1m_2m_3}b_{l_1l_2l_3}$ with $\mathcal{G}^{l_1l_2l_3}_{m_1m_2m_3}$ the Gaunt integral
\begin{equation}
\mathcal{G}^{l_1l_2l_3}_{m_1m_2m_3}=\int\mathrm{d}^2\hat{n}\,Y_{l_1m_1}(\hat{\vec{n}})Y_{l_2m_2}(\hat{\vec{n}})Y_{l_3m_3}(\hat{\vec{n}})\,.
\end{equation}
The optimal estimator in this case if given by
\begin{widetext}
\begin{equation}\label{eq:optbispecest}
\hat{f}_{\mathrm{NL}}^{\mathrm{opt}}:=\frac{1}{N}B^{l_1l_2l_3}_{m_1m_2m_3}(C^{+})_{l_1m_1l_4m_4}(C^{+})_{l_2m_2l_5m_5}(C^{+})_{l_3m_3l_6m_6}\left(\tilde{a}_{l_4m_4}\tilde{a}_{l_5m_5}\tilde{a}_{l_6m_6}-3\langle \tilde{a}_{l_4m_4}\tilde{a}_{l_5m_5}\rangle \tilde{a}_{l_6m_6}\right)\,,
\end{equation}
where $\mat{C}^+$ denotes the pseudoinverse of the covariance matrix $C_{l_1m_1l_2m_2}=\langle\tilde{a}_{l_1m_1}\tilde{a}^*_{l_2m_2}\rangle$ and $N$ is the normalisation factor
\begin{equation}
N=B^{l_1l_2l_3}_{m_1m_2m_3}(C^{+})_{l_1m_1l_4m_4}(C^{+})_{l_2m_2l_5m_5}(C^{+})_{l_3m_3l_6m_6}B^{l_4l_5l_6}_{m_4m_5m_6}\,.
\end{equation}
\end{widetext}
The estimator has variance
\begin{equation}
\mathrm{Var}[\hat{f}_{\mathrm{NL}}^{\mathrm{opt}}]=\frac{3!}{N}\,.
\end{equation}
The optimality property can be deduced from the Edgeworth expansion of the PDF in the connected $n$-point functions (see e.g.\ Refs.~\cite{Babich:OptEst, Taylor:Edgeworth} and references therein), which shows that the Fisher information at $f_{\mathrm{NL}}=0$ is indeed $F=N/3!$.

\subsection{Approximation to the optimal cut-sky bispectrum estimator}
\label{subsec:bispecapprox}
As in the case of the power spectra, optimal estimation of bispectra naively requires the inversion of the covariance matrix -- a computationally very challenging task. Even though full inverse-covariance weighting has been performed elsewhere (see e.g.\ \cite{Smith:OptimalLimits,Elsner:FastWienerFilter}) approximations are often used in practice to arrive at tractable estimators. In this work the estimator
\begin{widetext}
\begin{equation}\label{eq:bispecapproxapo}
\hat{f}_{\mathrm{NL}}=\frac{1}{\tilde{N}}\frac{B^{l_1l_2l_3}_{m_1m_2m_3}\left(\tilde{a}_{l_1 m_1}\tilde{a}_{l_2m_2}\tilde{a}_{l_3m_3}-3\langle \tilde{a}_{l_1m_1}\tilde{a}_{l_2m_2}\rangle\tilde{a}_{l_3m_3}\right)}{\bar{C}_{l_1}\bar{C}_{l_2}\bar{C}_{l_3}}
\end{equation}
\end{widetext}
is used, where $\tilde{N}$ is a suitably chosen normalisation ensuring unbiasedness,
\begin{equation}\label{eq:bispecapproxnorm}
\tilde{N}=\frac{B^{l_1l_2l_3}_{m_1m_2m_3}\tilde{B}^{l_1l_2l_3}_{m_1m_2m_3}}{\bar{C}_{l_1}\bar{C}_{l_2}\bar{C}_{l_3}}\,,
\end{equation}
and $\bar{C}_l=\langle\tilde{C}_l\rangle/f^{(2)}_{\mathrm{sky}}$ is the modified power spectrum introduced earlier in the context of covariance approximations. The appearance of the $\bar{C}_l$-factors as the weight can be understood as replacing the inverse of the covariance matrix by the inverse of its diagonal $C_{lmlm}\sim \langle\tilde{C_l}\rangle$. The choice to normalise with $f^{(2)}_{\mathrm{sky}}$ is not necessary because it could equally be absorbed into $\tilde{N}$ but convenient as it gives $\tilde{N}$ the familiar scaling $\propto f_{\mathrm{sky}}$. As discussed above, the mask function $U(\hat{\vec{n}})$ can be apodised to reduce long-range coupling of multipoles.

The approach is in a sense analogous to PCL estimators in the case of the power spectrum. In particular, if we assume that the power spectrum is constant and we are dealing with an unapodised mask, then this approximation to the estimator is exact. This can be simply seen by recalling that the covariance matrix is then proportional to a projection operator and is its own pseudoinverse and also $\bar{C}_l=C_l$ in this case. If both the power spectrum and the reduced bispectrum are constant, $b_{l_1l_2l_3}=b_0$, the normalisation factor is exactly given by
\begin{equation}\label{eq:fskyapproxtildeN}
\tilde{N}=f^{(3)}_{\mathrm{sky}}N_{\mathrm{fs}}\,,
\end{equation}
with $N_{\mathrm{fs}}$ denoting the full-sky normalisation
\begin{equation}
N_{\mathrm{fs}}=\sum_{l_im_i}\frac{(B^{l_1l_2l_3}_{m_1m_2m_3})^2}{C_{l_1}C_{l_2}C_{l_3}}\,.
\end{equation}
Similarly, the normalisation factor $N$ of the optimal estimator evaluates to
\begin{equation}
N=f^u_{\mathrm{sky}}N_{\mathrm{fs}}
\end{equation}
in this case. Even though they only strictly hold for constant underlying spectra, these results for the normalisation factors are simple to evaluate and useful in practice. We will refer to them as $f_{\mathrm{sky}}$-approximations to the normalisation factors.

\subsection{Leakage contributions to the variance}
\label{subsec:leakagecontrib}
The variance of the estimator is given by
\begin{widetext}
\begin{equation}\label{eq:bispecapproxvar}
\mathrm{Var}[\hat{f}_{\mathrm{NL}}]=\frac{6}{\tilde{N}^2}\frac{B^{l_1l_2l_3}_{m_1m_2m_3}}{(\bar{C}_{l_1}\bar{C}_{l_2}\bar{C}_{l_3})^{\frac{1}{2}}}\mathscr{C}_{l_1m_1l_4m_4}\mathscr{C}_{l_2m_2l_5m_5}\mathscr{C}_{l_3m_3l_6m_6}\frac{B^{l_4l_5l_6}_{m_4m_5m_6}}{(\bar{C}_{l_4}\bar{C}_{l_5}\bar{C}_{l_6})^{\frac{1}{2}}}\,,
\end{equation}
\end{widetext}
where we introduced the normalised covariance matrix
\begin{equation}\label{eq:normcovdef}
\mathscr{C}_{l_1m_1l_2m_2}=\frac{C_{l_1m_1l_2m_2}}{(\bar{C}_{l_1}\bar{C}_{l_2})^{\frac{1}{2}}}\,.
\end{equation}
Note that for the diagonal entries of this matrix we have
\begin{equation}\label{eq:normcovdiag}
\frac{1}{2l+1}\sum\limits_m\mathscr{C}_{lmlm}=f^{(2)}_{\mathrm{sky}}\,,
\end{equation}
independent of the shape of the power spectrum. With Eqs.~\eqref{eq:bispecapproxvar} and \eqref{eq:bispecapproxnorm} we can identify two phenomena related to mode coupling that affect the variance of PCL estimates. An increase in $\bar{C}_l$ due to leakage, as was particularly evident in the case of unapodised PCL estimation at high $l$ in Sec.~\ref{subsec:PCLandApo}, can cause $\tilde{N}$ in Eq.~\eqref{eq:bispecapproxnorm} to decrease. Even though the numerator in Eq.~\eqref{eq:bispecapproxvar} also decreases the $1/\tilde{N}^2$ factor should dominate and lead to a net increase in variance.

The second, and potentially more severe, effect can also be understood upon inspection of Eq.~\eqref{eq:bispecapproxvar}. Despite the fact that Eq.~\eqref{eq:normcovdiag} suggests that the normalised covariance in Eq.~\eqref{eq:bispecapproxvar} should not be directly affected by changes in $\bar{C}_l$, the variance $\mathrm{Var}[\hat{f}_{\mathrm{NL}}]$ can be extremely sensitive to mode coupling. The normalised covariance matrix is proportional to a projection operator in the case of a constant power spectrum and an unapodised mask, but it can develop very large eigenvalues if the power spectrum decays rapidly making it different from a simple projection as discussed in Ref.~\cite{Gruetjen:TowardsEfficient}. In particular, the strong initial decay of the power spectrum can generate large eigenvalues with eigenvectors related to the masked low-$l$ spherical harmonics. Inverse-covariance weighting would downweight the corresponding modes in the data and eliminate their effect on the variance. However, we see that the approximated estimator sums over the off-diagonal entries of the covariances, i.e.\ the correlated errors introduced by mode coupling, and can be heavily affected. To what extent obviously depends on the shape of the bispectrum under consideration.
\begin{figure*}[!tb]
\centering
\includegraphics[width=.48\textwidth]{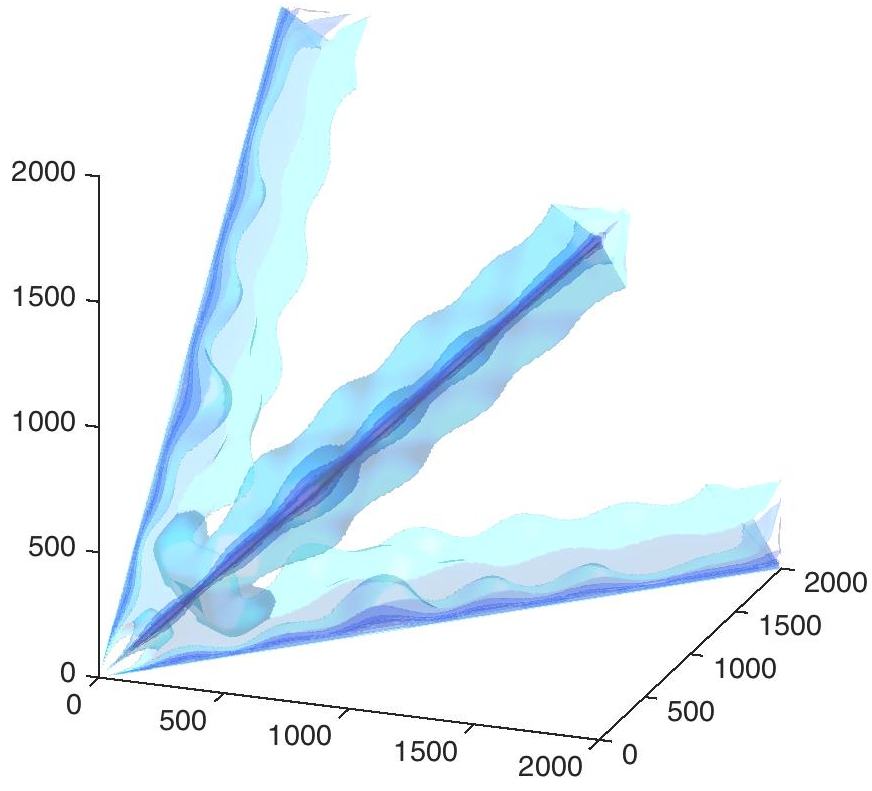}
\includegraphics[width=.48\textwidth]{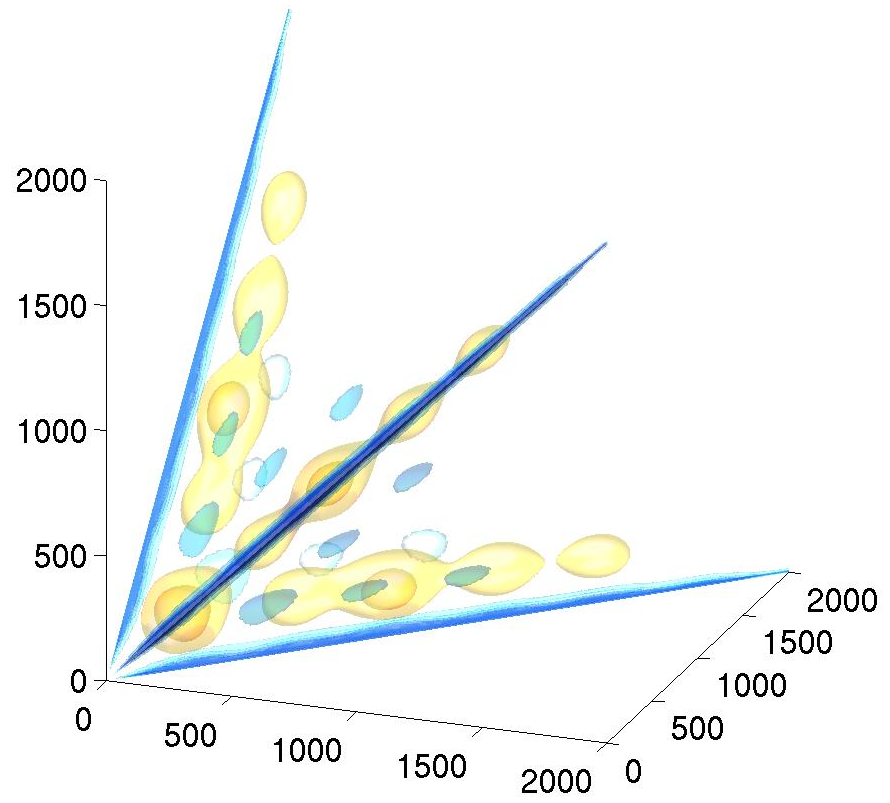}
\caption{Plot of the reduced bispectra of the mask (left) and the local shape (right) based on a modal reconstruction and weighted with the large-angle solution for the constant shape (see main text). Note that the modal reconstruction is only able to capture a small component of the full, highly oscillatory mask bispectrum. The reconstruction is based on relatively slowly varying basis functions that describe physical reduced bispectra and thus ideally suited for visualising the components of the mask bispectrum that are relevant for constraining cosmology.}
\label{fig:bispecmasklocal}
\end{figure*}
To make this more explicit we use the results from Ref.~\cite{Gruetjen:TowardsEfficient} mentioned above to estimate the contribution to the variance from these large eigenvalues. In particular, assuming a simple symmetric galactic cut there are only two of these eigenvectors for each $m$. One corresponding to even parity and one corresponding to odd parity. Let us focus on the even parity eigenvector associated with $m=0$ and label it as $\vec{v}_0$. It is approximately given by%
\footnote{We ignore small differences between $\bar{C}_l$ and $C_l$ here as they do not affect the conclusions.}
\begin{equation}
(\vec{v}_0)_{lm}=\frac{1}{n_v}\frac{u_{lm}}{C_l^{\frac{1}{2}}}\,,
\end{equation}
i.e.\ proportional to the multipole coefficients of the unapodised mask function $U(\hat{\vec{n}})$ itself. Here, $n_v$ is a normalisation factor,
\begin{equation}
n_v^2=\sum_{lm}\frac{\vert u_{lm}\vert^2}{C_l}\,,
\end{equation}
chosen to make $\vec{v}_0$ a unit vector. Let us call the corresponding eigenvalue of the normalised covariance $\lambda$. This eigenvector generates a contribution to the normalised covariance
\begin{equation}
\mathscr{C}_{l_1m_1l_2m_2}\sim \lambda(\vec{v}_0)_{l_1m_1}(\vec{v}_0)^*_{l_2m_2}=\frac{\lambda}{n_v^2}\frac{u_{l_1m_1}}{C_{l_1}^{\frac{1}{2}}}\frac{u^*_{l_2m_2}}{C_{l_2}^{\frac{1}{2}}}
\end{equation}
and the contribution to $\mathrm{Var}[\hat{f}_{\mathrm{NL}}]$ can be estimated to be
\begin{align}\nonumber
\frac{\mathrm{Var}[\hat{f}_{\mathrm{NL}}]}{\mathrm{Var}[\hat{f}^{\mathrm{opt}}_{\mathrm{NL}}]}&\sim\frac{\lambda^3}{\tilde{N}n_v^6}\left(\sum_{l_im_i}\frac{u_{l_1m_1}u_{l_2m_2}u_{l_3m_3}B^{l_1l_2l_3}_{m_1m_2m_3}}{C_{l_1}C_{l_2}C_{l_2}}\right)^2\\\nonumber
&=\frac{\lambda^3}{\tilde{N}n_v^6}\left(\sum_{l_im_i}\frac{U^{l_1l_2l_3}_{m_1m_2m_3}B^{l_1l_2l_3}_{m_1m_2m_3}}{C_{l_1}C_{l_2}C_{l_2}}\right)^2\\\label{eq:vardegradest}
&=\frac{\lambda^3}{f^u_{\mathrm{sky}}}\mathrm{CorrB}[B,U]^2\,,
\end{align}
where we introduced the mask bispectrum
\begin{equation}
U^{l_1l_2l_3}_{m_1m_2m_3}=u_{l_1m_1}u_{l_2m_2}u_{l_3m_3}
\end{equation}
and the bispectrum correlator
\begin{widetext}
\begin{align}
\mathrm{CorrB}[X,Y]&=\frac{1}{(N_{\mathrm{fs}}(X)N_{\mathrm{fs}}(Y))^{\frac{1}{2}}}\sum_{l_im_i}\frac{X^{l_1l_2l_3}_{m_1m_2m_3}Y^{l_1l_2l_3}_{m_1m_2m_3}}{C_{l_1}C_{l_2}C_{l_2}}\\
&=\frac{1}{(N_{\mathrm{fs}}(X)N_{\mathrm{fs}}(Y))^{\frac{1}{2}}}\sum_{l_i}\frac{(2l_1+1)(2l_2+1)(2l_3+1)}{4\pi}\left(\begin{array}{ccc} l_1 & l_2 & l_3 \\ 0 & 0 & 0 \end{array}\right)^{2}\frac{x_{l_1l_2l_3}y_{l_1l_2l_3}}{C_{l_1}C_{l_2}C_{l_2}}\,,
\end{align}
\end{widetext}
where the expression involving the reduced bispectra $x_{l_1l_2l_3}$ and $y_{l_1l_2l_3}$ holds if at least one of the bispectra is statistically isotropic. As discussed in Ref.~\cite{Gruetjen:TowardsEfficient} one can think of the normalised covariance matrix in terms of oblique projections giving rise to a simple geometrical picture that can be used to estimate the order of magnitude of $\lambda$ for an unapodised mask. We provide the details of the calculation in App.~\ref{app:mageigen}. For the setup studied here, the result is $\lambda^3\approx 4\times 10^{13}$. Hence, we can estimate the contribution from this eigenvalue to the variance as
\begin{equation}
\frac{\mathrm{Var}[\hat{f}_{\mathrm{NL}}]}{\mathrm{Var}[\hat{f}^{\mathrm{opt}}_{\mathrm{NL}}]}\sim 5\times10^{13}\,\mathrm{CorrB}[B,U]^2\,.
\end{equation}
The $u_{lm}$ obviously do not all have the same sign but rather oscillate rapidly, so we also expect the mask bispectrum $U^{l_1l_2l_3}_{m_1m_2m_3}$ and the associated reduced bispectrum%
\footnote{Note that the definition of $u_{l_1l_2l_3}$ only applies for even $l_1+l_2+l_3$ (otherwise we would be dividing by zero). As we are typically interested in correlators involving cosmological isotropic bispectra only this case matters and there is no need to define $u_{l_1l_2l_3}$ for odd $l_1+l_2+l_3$.} %
\begin{widetext}
\begin{equation}
u_{l_1l_2l_3}=\sqrt{\frac{4\pi}{(2l_1+1)(2l_2+1)(2l_3+1)}}\left(\begin{array}{ccc} l_1 & l_2 & l_3 \\ 0 & 0 & 0 \end{array}\right)^{-1}\sum_{m_i}\left(\begin{array}{ccc} l_1 & l_2 & l_3 \\ m_1 & m_2 & m_3 \end{array}\right)U^{l_1l_2l_3}_{m_1m_2m_3} 
\end{equation}
\end{widetext}
to oscillate rapidly. The oscillatory behaviour of the reduced bispectrum can be seen particularly well under the assumption of a simple symmetric galactic cut. In this case only the $u_{l0}$ with even $l$ are nonzero, the Wigner $3j$ symbols cancel and thus $u_{l_1l_2l_3}\propto u_{l_10}u_{l_20}u_{l_30}$. Thus, the reduced bispectrum inherits the highly oscillatory character of the mask multipole coefficients. This means that we generally only expect small correlations with shapes of cosmological interest that typically have a relatively slowly varying reduced bispectrum. Using $u_{lm}\sim l^{-1}$ to roughly describe the scaling of the mask multipole coefficients, the overall scaling of the reduced mask bispectrum is $u_{l_1l_2l_3}\sim(l_1l_2l_3)^{-3/2}$, which suggest a squeezed character. Hence, we expect the mask bispectrum to be highly oscillatory with a scaling that resembles that of local-type non-Gaussianity.

Fig.~\ref{fig:bispecmasklocal} plots the reduced bispectra of the mask and the local shape obtained using a modal reconstruction \cite{Fergusson:PrimordialNonGaussianity}. The reduced bispectra are weighted with the large-angle solution for the reduced bispectrum of the constant shape presented in Ref.~\cite{Fergusson:shapeofNG},
\begin{equation}
b^{\mathrm{const}}_{l_1l_2l_3}\propto\left(\prod\limits_i\frac{1}{2l_i+1}\right)\left(\frac{1}{l_1+l_2+l_3+3}+\frac{1}{l_1+l_2+l_3}\right)\,.
\end{equation}
The modal reconstruction only uses a limited set of slowly varying modes relevant to cosmology to approximate the reduced bispectra and thus is blind to highly oscillatory behaviour. However, the plot clearly shows how the modal reconstruction of $u_{l_1l_2l_3}$ indeed has a squeezed character and looks similar to the local shape suggesting a relatively large correlation $\mathrm{CorrB}[B^{\mathrm{loc}},U]$. Precise values for the correlation $\mathrm{CorrB}[B,U]$ of the mask bispectrum with various standard shapes are listed in Table~\ref{tab:maskcorrs}.
\begin{table}[tb]
\centering
\caption{Values for the correlation $\mathrm{CorrB}[B,U]$ of the mask bispectrum $U^{l_1l_2l_3}_{m_1m_2m_3}$ with various standard shapes.}
\begin{tabular}{@{} lcc @{}}
    \hline
    Shape        &  \makebox[2.5cm][c]{$\mathrm{CorrB}[B,U]$} & $5\times 10^{13}\,\mathrm{CorrB}[B,U]^2$\\\hline
    local        &  $2.4\times 10^{-6}$  & $\mathcal{O}(10^2)$\\ 
    equilateral  &  $-4.8\times 10^{-7}$ & $\mathcal{O}(10^1)$\\
    flat         &  $3.8\times 10^{-7}$  & $\mathcal{O}(10^1)$\\  
    constant     &  $-4.8\times 10^{-8}$ & $\mathcal{O}(10^{-1})$\\
    \hline
\end{tabular}
\label{tab:maskcorrs}
\end{table}
As expected, all correlations are very small due to the oscillatory nature of the mask bispectrum but the correlation with the local shape is by far the largest. It is five times larger than the correlation with the equilateral and flat shape and almost two orders of magnitude larger than the correlation with the constant shape. Table~\ref{tab:maskcorrs} also presents estimates for the contribution to the variance. We expect a very significant contamination of local measurements increasing the variance by orders of magnitude and so we will focus on the local shape to numerically investigate the effect of leakage on bispectrum estimates.

Having discussed the relevance of mode coupling for bispectrum measurements, we will briefly discuss various approaches to reducing leakage. As in the case of the power spectrum we can obviously apodise the mask. This will significantly reduce long-range leakage and should lower the variance. However, it inevitably results in a loss of sky fraction and, unless we go to extreme apodisation scales, there will still be some leakage contributing to $\bar{C}_l$, both of which can degrade the errors. We will investigate the performance of apodised bispectrum estimates in Sec.~\ref{sec:localresults}.

\subsection{Reducing leakage by explicit subtraction} 
Another approach to reducing leakage that we implement is a cleaning scheme that systematically subtracts leakage from low-$l$ multipoles at the $\tilde{a}_{lm}$ level. Specifically, starting with a given set of $\tilde{a}_{lm}$ we can subtract out all low-$l$ contributions to high-$l$ $\tilde{a}_{lm}$ by finding the minimum norm solution $m_{lm}$ to the equation
\begin{equation}
P_{l_1m_1l_2m_2}m_{l_2m_2}=\tilde{a}_{l_1m_1}
\end{equation}
working up to a given cleaning scale $l_{\mathrm{cl}}$, i.e.\ $\mat{P}$ is a $(l_{\mathrm{cl}}+1)^2\times(l_{\mathrm{cl}}+1)^2$ matrix in this equation. The cleaned $\tilde{a}^c_{lm}$ are then chosen to be
\begin{equation}
\tilde{a}^c_{lm}=\begin{cases}\tilde{a}_{lm} & \mbox{for } l\le l_{\mathrm{cl}}+20 \\ \tilde{a}_{lm}-P_{lml_1m_1}m_{l_1m_1}& \mbox{for } l>l_{\mathrm{cl}}+20 \end{cases}\,,
\end{equation}
where the transition from unchanged $\tilde{a}_{lm}$ to low-$l$ subtracted $\tilde{a}_{lm}$ at $l_{\mathrm{cl}}+20$ is chosen so that we avoid those low-$l$ subtracted multipoles just beyond $l_{\mathrm{cl}}$ that have a significant fraction of their power subtracted due to strong coupling to $\tilde{a}_{lm}$ with $l\le l_{\mathrm{cl}}$. This approach has the advantage that it eliminates low-high coupling originating from the low-$l$ multipoles without losing sky coverage. The cleaning is evidently limited to low multipoles and leakage originating from higher multipoles is not reduced. In particular at high $l$, the resulting $\bar{C}_l$ will not change significantly as only a fraction of the leakage contribution to $\bar{C}_l$ arises from the low-$l$ multipoles. However, the fact that it completely eliminates all leakage from low-$l$ multipoles makes it an interesting approach to study as it highlights the problems that solely arise from this type of leakage. Note that it can be thought of as an extreme downweighting of the modes corresponding to masked low-$l$ spherical harmonics that completely eliminates them. This is an approximation to what we would expect inverse-covariance weighting of the $\tilde{a}_{lm}$ to achieve, given that these are the modes that correspond to the large eigenvalues of the normalised covariance as discussed above. Cleaning and apodisation can also be combined, i.e.\ cleaning can be carried out on apodised $\tilde{a}_{lm}$ to quantify the impact of residual low-$l$ contamination in the apodised cases.

\subsection{Bispectrum estimates from inpainted maps}
An alternative to apodisation and explicit cleaning is inpainting. First of all note that inpainting does not introduce a bias of the estimator in the form of a spurious bispectrum because it is linear in the $a_{lm}$ as discussed in Sec.~\ref{subsec:InpPS}. Hence, Gaussian $a_{lm}$ will give rise to Gaussian $\tilde{a}^{\mathrm{I}}_{lm}$ and deviations from Gaussianity can still be estimates with the optimal bispectrum estimator Eq.~\eqref{eq:optbispecest}. However, there is one caveat. The full-sky bispectrum $B^{l_1l_2l_3}_{m_1m_2m_3}$ appeared in Eq.~\eqref{eq:optbispecest} rather than the cut-sky bispectrum $\tilde{B}^{l_1l_2l_3}_{m_1m_2m_3}$. This was correct because full-sky and cut-sky bispectrum were simply related by projection operators $\mat{P}$ that leave $\mat{C}^+$ unchanged so that there is no need to work with the cut-sky bispectrum. This is not true for the inpainted bispectrum $(\tilde{B}^{\mathrm{I}})^{l_1l_2l_3}_{m_1m_2m_3}$ and we need to take this into account when constructing an approximation to the optimal estimator based on inpainted maps $\hat{f}^{\mathrm{I}}	$. 

Apart from this detail, the approximation to the optimal estimator is formally identical and reads
\begin{widetext}
\begin{equation}
\hat{f}^{\mathrm{I}}_{\mathrm{NL}}=\frac{1}{\tilde{N}}\frac{(\tilde{B}^{\mathrm{I}})^{l_1l_2l_3}_{m_1m_2m_3}\left(\tilde{a}^{\mathrm{I}}_{l_1 m_1}\tilde{a}^{\mathrm{I}}_{l_2m_2}\tilde{a}^{\mathrm{I}}_{l_3m_3}-3\langle \tilde{a}^{\mathrm{I}}_{l_1m_1}\tilde{a}^{\mathrm{I}}_{l_2m_2}\rangle\tilde{a}^{\mathrm{I}}_{l_3m_3}\right)}{\bar{C}_{l_1}\bar{C}_{l_2}\bar{C}_{l_3}}\,.
\end{equation}
Now, recalling the discussion above about the amplification of $a_{lm}$ by inpainting at low $l$ described by the factor $\Pi^{\mathrm{I}}_{ll}/\Pi_{ll}$, a natural replacement is
\begin{equation}
\frac{(\tilde{B}^{\mathrm{I}})^{l_1l_2l_3}_{m_1m_2m_3}}{(\bar{C}_{l_1}\bar{C}_{l_2}\bar{C}_{l_3})^{\frac{1}{2}}}\rightarrow\frac{B^{l_1l_2l_3}_{m_1m_2m_3}}{(\bar{C}_{l_1}\bar{C}_{l_2}\bar{C}_{l_3})^{\frac{1}{2}}}\frac{(\Pi^{\mathrm{I}}_{l_1l_1}\Pi^{\mathrm{I}}_{l_2l_2}\Pi^{\mathrm{I}}_{l_3l_3})^{\frac{1}{2}}}{(\Pi_{l_1l_1}\Pi_{l_2l_2}\Pi_{l_3l_3})^{\frac{1}{2}}}=\frac{B^{l_1l_2l_3}_{m_1m_2m_3}}{(\bar{C}^{\mathrm{I}}_{l_1}\bar{C}^{\mathrm{I}}_{l_2}\bar{C}^{\mathrm{I}}_{l_3})^{\frac{1}{2}}}
\end{equation}
The approximation then reads
\begin{equation}\label{eq:bispecapproxinp}
\hat{f}^{\mathrm{I}}_{\mathrm{NL}}=\frac{1}{\tilde{N}}\sum\limits_{l_im_i}\frac{B^{l_1l_2l_3}_{m_1m_2m_3}}{(\bar{C}^{\mathrm{I}}_{l_1}\bar{C}^{\mathrm{I}}_{l_2}\bar{C}^{\mathrm{I}}_{l_3})^{\frac{1}{2}}}\frac{\left(\tilde{a}^{\mathrm{I}}_{l_1 m_1}\tilde{a}^{\mathrm{I}}_{l_2m_2}\tilde{a}^{\mathrm{I}}_{l_3m_3}-3C^{\mathrm{I}}_{l_1m_1l_2m_2}\tilde{a}^{\mathrm{I}}_{l_3m_3}\right)}{(\bar{C}_{l_1}\bar{C}_{l_2}\bar{C}_{l_3})^{\frac{1}{2}}}\,.
\end{equation}
\end{widetext}
We saw in the previous sections that inpainting is extremely efficient at reducing low-high leakage without suffering from a loss of sky coverage. So we expect it to completely eliminate any effect of low-$l$ leakage into high $l$ modes just as in the case of the cleaning scheme described above. However, it has the further advantage that it eliminates the leakage effects originating at higher $l$ as well. This results in a $\bar{C}^{\mathrm{I}}_l$ for inpainting that is close to $C_l$. Recalling the discussion of contributions to the estimator variance in Sec.~\ref{subsec:leakagecontrib} one can hope that these effects taken together enable inpainting to prevent any leakage induced suboptimalities.

\subsection{Numerical results for the local shape}
\label{sec:localresults}
To check our analytic expectations from the previous sections numerically and compare different approaches to reducing leakage, we obtain estimator variances for the local shape from MC simulations. In principle, we should be evaluating Eq.~\eqref{eq:bispecapproxapo} for the apodisation and cleaning cases and Eq.~\eqref{eq:bispecapproxinp} for the inpainting case. To evaluate Eq.~\eqref{eq:bispecapproxinp} knowledge of $\bar{C}^{\mathrm{I}}_l$ is required, which in turn necessitates the calculation of $\mat{\Pi}^{\mathrm{I}}$. Rather than calculating $\mat{\Pi}^{\mathrm{I}}$ at HEALPix resolution $N_{\mathrm{side}}=2048$ up to $l_{\mathrm{max}}=2000$, we simply approximate $\bar{C}^{\mathrm{I}}_l\approx C_l$ (a rather good approximation according to Fig.~\ref{fig:barCl}). To place all methods on the same footing, we make the same replacement in the apodisation and cleaning cases as well, i.e.\ we evaluate the estimator
\begin{widetext}
\begin{equation}\label{eq:bispecapproxprac}
\hat{f}^{\mathrm{I}}_{\mathrm{NL}}=\frac{1}{\tilde{N}}\frac{B^{l_1l_2l_3}_{m_1m_2m_3}}{(C_{l_1}C_{l_2}C_{l_3})^{\frac{1}{2}}}\frac{\left(\tilde{a}_{l_1 m_1}\tilde{a}_{l_2m_2}\tilde{a}_{l_3m_3}-3C_{l_1m_1l_2m_2}\tilde{a}_{l_3m_3}\right)}{(\bar{C}_{l_1}\bar{C}_{l_2}\bar{C}_{l_3})^{\frac{1}{2}}}
\end{equation}
\end{widetext}
for all methods, where $\tilde{a}_{lm}$ and $\bar{C}_l=\langle\tilde{C}_l\rangle/f^{(2)}_{\mathrm{sky}}$ are either the pseudomultipoles and pseudospectra obtained from apodisation (and cleaning%
\footnote{Cleaning has a subtle impact on the pseudospectra as well. This is because due to the coupling of the higher-$l$ multipoles to the low-$l$ $a_{lm}$, some of the power is subtracted out. We take it into account here, despite the fact that this only leads to very small changes in the weightings and is unlikely to have any significant effect on estimator variances.}) %
or those obtained from inpainting%
\footnote{Because we did not calculate $\mat{\Pi}^{\mathrm{I}}$, we obtained the inpainting pseudospectrum $\langle\tilde{C}^{\mathrm{I}}_l\rangle$ from $10^4$ MC samples, which offers more than enough accuracy for the purposes of determining a weighting factor.}. %
We checked explicitly for some of the apodisation cases, where $\bar{C}_l$ can generally be calculated easily, that the subtle difference in weighting has negligible effect on the resulting variance. Potential differences are very small and within the error bars on the variances so that this does not affect any of the conclusions.
\begin{table*}[tb]
\centering
\caption{Error bars for the local estimator, $\Delta f_{\mathrm{NL}}=\mathrm{Var}[\hat{f}_{\mathrm{NL}}]^{\frac{1}{2}}$, for various combinations of apodisation and cleaning as well as inpainting. Results are obtained from either $200$ or $800$ realisations. The latter case is indicated with an asterisk. The error bars are obtained using exact normalisation factors as obtained from MC sampling using simulated non-Gaussian maps. Uncertainties in the error bars include those arising from MC errors on the normalisation factors. The $f_{\mathrm{sky}}$-approximation to the optimal limit is $\Delta f_{\mathrm{NL}}=(3!/(f^{\mathrm{u}}_{\mathrm{sky}}N_{\mathrm{fs}}))^{\frac{1}{2}}\approx 5.84$.}
\begin{tabular}{@{} lcccc @{}}
    \hline
                            &  \makebox[2cm][c]{Unapodised} & \makebox[2cm][c]{S015} & \makebox[2cm][c]{S03} & \makebox[2cm][c]{S05} \\\hline
    Uncleaned  &  $31.53\,(\pm 2.89)$ & $13.37\,(\pm 0.80)$ &$7.40\,(\pm 0.23)^*$& $7.16\,(\pm 0.22)^*$\\ 
    Cleaned $l_{\mathrm{cl}}=30$ &  $10.07\,(\pm 0.56)$ & $6.41 \,(\pm 0.34)$ & $6.00\,(\pm 0.17)^*$ & $6.27\,(\pm 0.33)$\\
    Cleaned $l_{\mathrm{cl}}=70$ &  $9.51 \,(\pm 0.52)$ & $6.39 \,(\pm 0.19)^*$ & $6.00\,(\pm 0.17)^*$& $6.54\,(\pm 0.19)^*$\\  
    \hline\hline
                            & Inpainting     & $5.88\,(\pm 0.17)^*$\\
    \hline
\end{tabular}
\label{tab:localvars}
\end{table*}

Table~\ref{tab:localvars} lists the standard deviations $\Delta f_{\mathrm{NL}}=\mathrm{Var}[\hat{f}_{\mathrm{NL}}]^{\frac{1}{2}}$ of local estimates obtained for the various cases from either $200$ or $800$ Gaussian simulations. The larger number of $800$ simulations was used to obtain more accurate error bars in certain cases marked with an asterisk in Table~\ref{tab:localvars}. In the apodisation and cleaning cases, we employ the unapodised, S015 and S05 mask and provide results for no cleaning and cleaning up to $l_{\mathrm{cl}}=30$ and $l_{\mathrm{cl}}=70$. The normalisation factors $\tilde{N}$ in Eq.~\eqref{eq:bispecapproxprac} were calculated exactly (up to MC errors) from 420 non-Gaussian local simulations ($f_{\mathrm{NL}}=20$) to avoid potential inaccuracies of the $f_{\mathrm{sky}}$-approximation to the normalisation factors affecting the results. For most of the cases, the residual MC errors on the normalisation factors are well below the uncertainty in the variances using $200$ maps (but are included in the quoted uncertainties nonetheless). 

First of all, we observe highly suboptimal error bars for the unapodised estimates without any cleaning. The $f_{\mathrm{sky}}$-approximation for the optimal error bar based on $N=f^{\mathrm{u}}_{\mathrm{sky}}N_{\mathrm{fs}}$ predicts a standard deviation $\Delta f\approx 5.84$. Thus, the unapodised and uncleaned case with $\Delta f\approx 31.5$ produces a variance roughly $30$ times larger than the optimal limit. The analytic considerations in the last section produced a rough estimate $\mathcal{O}(10^2)$ for this ratio. Given the approximate nature of the estimate, the agreement is reasonably good. Subtracting the low-$l$ power leads to enormous improvements reducing the variance by more than a factor of 10. Here, most improvements come from the lowest $l$ with little further gains made by increasing $l_{\mathrm{cl}}$ from 30 to 70. This behaviour confirms the picture discussed in the last sections that the contributions due to correlated errors are mainly driven by the low-$l$ multipoles. However, cleaning alone is not enough and we do not arrive at near-optimal estimates even with $l_{\mathrm{cl}}=70$, where we find $\Delta f_{\mathrm{NL}}\approx 9.5$. We can attribute this to the significant leakage contributions to $\bar{C}_l$ in the unapodised case that also originate at higher $l$ and are thus still partially present in the cleaned scenarios.

Focusing next on the apodised cases without any cleaning, we see that an apodisation scale of $0.15^{\circ}$ still leads to substantial suboptimality. In contrast the S05 mask produces error bars that are already only $20\%$ above optimal without cleaning. The suppression of long-range leakage due to apodisation both reduces leakage contributions to $\bar{C}_l$ and also ameliorates the impact of correlated errors from low-$l$ multipoles. The arguments that gave rise to an estimate of the large eigenvalues of the normalised covariance only directly apply to the case of unapodised masks. However, we argue in App.~\ref{app:mageigen} that the method should also provide a rough idea of the magnitude of $\lambda$ in the apodised cases giving for example $\lambda_{\mathrm{S015}}^3=\mathcal{O}(10^{12})$ and $\lambda_{\mathrm{S05}}^3=\mathcal{O}(10^{10})$. Using the estimate that relates $\lambda$ to the degradation of the variance, Eq.~\eqref{eq:vardegradest}, this translates to relative contributions to the variance of $\mathcal{O}(10)$ and $\mathcal{O}(0.1)$, respectively. This is in reasonable agreement with the actual results showing a more than fivefold increased variance in the S015 case compared to the optimal limit and a roughly 50 percent increased variance in the S05 case.

Subtraction of the low-$l$ power in conjunction with apodisation leads to further improvements. After subtraction of the low-$l$ power the error bars are as low as $\Delta f_{\mathrm{NL}}\approx 6.0$ in the case of the S03 mask. This is slightly larger than the $f_{\mathrm{sky}}$-approximation to the optimal limit $5.84$ which is just within the one sigma uncertainty of the measured error bar. More simulations would be needed to produce clear evidence that the S03 mask in conjunction with low-$l$ cleaning is still suboptimal.

\begin{figure*}[tb]
\centering
\includegraphics[width=.7\textwidth]{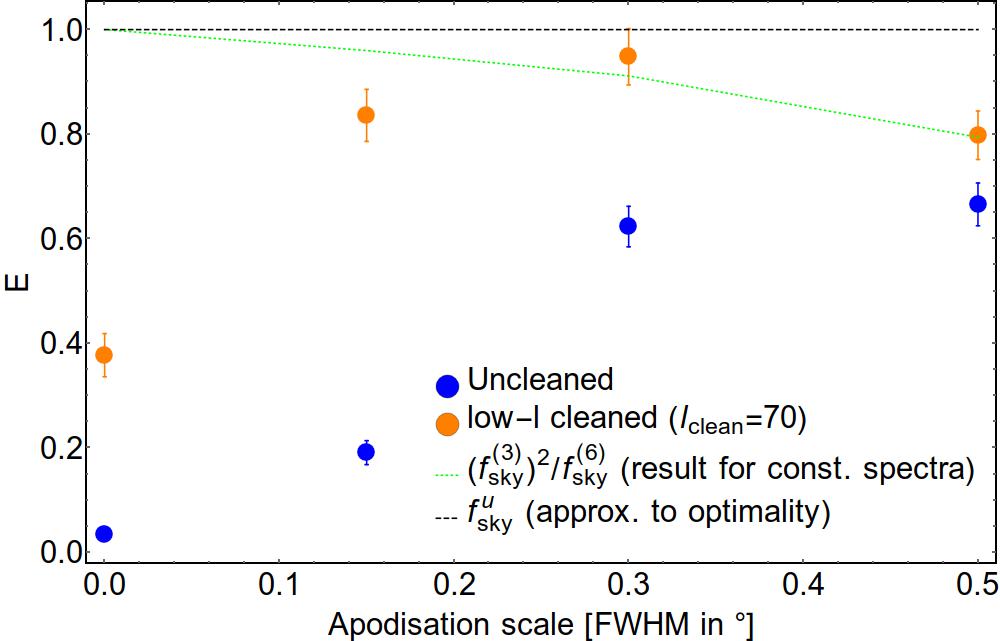}
\caption{The efficiency of the local estimator for the different apodisation scales in the $l_{\mathrm{cl}}=70$ (orange) and uncleaned (blue) case. Also shown is the prediction based on the assumption of constant spectra (green dotted line) and the optimal limit (dashed black line).}
\label{fig:SNplotbispec}
\end{figure*}

Note that the low-$l$ cleaned error bars have a minimum at an apodisation scale of $0.3^{\circ}$ and begin to grow again for more aggressive apodisation. To obtain further insight the apodisation and cleaning results of this section are visualised in Fig.~\ref{fig:SNplotbispec}.
The figure plots the efficiency, $E[\hat{f}_{\mathrm{NL}}]$, of the local estimators for the different apodisation scales in the $l_{\mathrm{cl}}=70$ and uncleaned case. The efficiency of the estimator is defined as the ratio of the optimal variance and the estimator variance and given by
\begin{equation}
E[\hat{f}_{\mathrm{NL}}]=\frac{1/F}{\mathrm{Var}[\hat{f}_{\mathrm{NL}}]}=\frac{3!}{f^{\mathrm{u}}_{\mathrm{sky}}\,N_{\mathrm{fs}}\mathrm{Var}[\hat{f}_{\mathrm{NL}}]}\,,
\end{equation}
where we made use of the $f_{\mathrm{sky}}$-approximation to the normalisation factor to calculate the Fisher information. The figure also includes a prediction based on the assumption of constant spectra. In this case we have
\begin{equation}
\mathrm{Var}[\hat{f}_{\mathrm{NL}}]=\frac{3!f^{(6)}_{\mathrm{sky}}}{(f^{(3)}_{\mathrm{sky}})^2N_{\mathrm{fs}}}\quad\Rightarrow\quad E[\hat{f}_{\mathrm{NL}}]=\frac{(f^{(3)}_{\mathrm{sky}})^2}{f^{(6)}_{\mathrm{sky}}f^{\mathrm{u}}_{\mathrm{sky}}}\,.
\end{equation}
For constant spectra, leakage is irrelevant and the decrease in efficiency is purely due to the loss of sky fraction resulting from the smoothing of the mask. In this sense this curve provides us with an expectation of how much the loss of sky fraction alone degrades the errors. Plotting the results in this way highlights how for small apodisation scales the leakage leads to estimator efficiencies far below what one would expect simply based on the loss of sky fraction. However, for sufficient apodisation the observed efficiencies start to agree with the constant spectra (i.e.\ no leakage) prediction. Cleaning the low-$l$ power mitigates the leakage induced suboptimalities by eliminating the variance contributions from the large eigenvalues of the normalised covariance and already leads to good agreement with the constant spectra approximation at an apodisation scale of $0.3^{\circ}$. Further apodisation beyond this point leads to a decrease in efficiency following the constant spectra approximation. This explains the minimum in the low-$l$ cleaned error bars already mentioned in the discussion of Table~\ref{tab:localvars}.

The results suggest that with low-$l$ cleaning a balance between reducing leakage and minimising the loss of sky fraction is struck at smaller apodisation scales than without low-$l$ cleaning and the maximum achievable efficiency is increased to $\sim 90\%$ compared to $\lesssim 70\%$ without low-$l$ cleaning.

Having discussed the apodisation and cleaning cases let us move on to inpainting. Inpainting produces an error bar $\Delta f_{\mathrm{NL}}\approx 5.9$. This agrees very well with the $f_{\mathrm{sky}}$-approximation of the optimal limit at the accuracy with which we measured estimator variances. Hence, assuming that the $f_{\mathrm{sky}}$-approximation is accurate, we conclude that inpainting is indeed optimal or at least very nearly optimal. From the point of view taken here this is because it heavily suppresses low-high leakage, thus eliminating any contributions due to the large eigenvalues of $\mat{\mathscr{C}}$ as well as additional leakage contributions to $\bar{C}_l^{\mathrm{I}}$ arising at higher $l$, and it does not suffer from a loss of sky fraction.

\section{Summary and discussion}
In this paper we discussed inpainting as an approach to constructing accurate CMB estimators and compared its performance to other methods. We first studied the case of the power spectrum and showed that it is possible to utilise the linearity of inpainting to construct analytically debiased power spectrum estimates from inpainted maps. The estimator significantly outperforms PCL estimates obtained from unapodised or apodised masks. We provided an explanation of this fact based on the observation that, in contrast to PCL, inpainting couples multipoles asymmetrically. While high-low coupling is exacerbated, inpainting is extremely effective at suppressing low-high coupling. This makes inpainting very suitable for the analysis of CMB temperature fluctuations that exhibit strongly decaying spectra so that low-high coupling is important while high-low coupling is largely irrelevant. Unlike apodisation of the mask, inpainting does not reduce the effective sky fraction so that this suppression of leakage does not come at the price of reduced sky coverage that leads to a degradation of errors. Comparing the variance of PCL estimates to estimates from inpainting, we observed improvements in variance exceeding $20\%$ depending on the apodisation scale used for the PCL estimates.

We also proposed improved approximations to PCL covariance matrices that allow for a generalisation to the covariance matrices of inpainted power spectrum estimates. The resulting approximations are in both cases highly accurate on the diagonal. Assuming that the off-diagonal agreement is sufficient, inpainting offers a viable and more accurate alternative to apodised PCL estimation as a framework for CMB power spectrum analysis and the construction of a likelihood.

We proceeded to study the case of approximations to the optimal cut-sky bispectrum estimator. We derived an analytic estimate for the contamination of bispectrum estimates for a given shape due to masking the data. The estimate relates the expected contribution to the variance to the bispectrum correlator between a given shape and a mask bispectrum. It provides an explanation of the empirical fact that the local shape is affected most, having by far the largest correlation with the mask bispectrum amongst the shapes of cosmological interest.

Focusing on the local shape we went on to study estimator variances numerically. We compared three techniques of ameliorating mode coupling: inpainting, apodisation and direct subtraction of power from coupling to low-$l$ modes. The results gave a variance for the unapodised and uncleaned case 30 times higher than the optimal limit, confirming the expectation from the analytic estimate. Cleaning out the low-$l$ contributions led to the expected significant improvements, but fell short of completely restoring optimality due to leakage contributions arising at higher $l$. Apodising the mask also had a very beneficial effect. Sufficient apodisation suppresses the long-range leakage substantially and gave rise to estimators that are relatively near, but still noticeably above, the optimal limit without any low-$l$ cleaning. 

Combining apodisation with cleaning of the low-$l$ modes further reduced the variances producing estimates that came very close in performance to the optimal limit. Our results for inpainted bispectrum estimates suggested that these still tend to slightly outperform any of the other methods. The measured variance agreed well with the $f_{\mathrm{sky}}$-approximation to the optimal limit. As in the case of the power spectrum, we attribute the efficacy of inpainting to the strong suppression of low-high leakage that does not come at the price of a loss of sky fraction. Even though we showed that it is possible to construct alternative numerically simple estimators with comparably small error bars in the case of the bispectrum, inpainting is probably still the preferable method due to its conceptual simplicity and straightforward implementation.

We conclude that there is no need to think of inpainting merely as a crude ad-hoc solution to curing suboptimalities of cut-sky bispectrum estimators. It is a highly effective approach to reducing leakage contributions to the variance of cut-sky estimators in general when the underlying spectra decay rapidly. Inpainting is a vital tool in bispectrum analysis already and could prove very useful for power spectrum analysis and elsewhere.

We did not explore inpainting in the context of the analysis of CMB polarisation data and leave this to future work. The polarisation spectra do not decay rapidly like the temperature spectra and different conclusions concerning the performance of inpainting might be reached.

\begin{acknowledgments} 
We thank Anthony Challinor, Carlo Contaldi and Benjamin Wallisch for very useful discussions. We are grateful to Juha J\"{a}ykk\"{a} and James Briggs for outstanding computational support. HFG gratefully acknowledges the support of the Studienstiftung des deutschen Volkes, an STFC studentship and a studentship of the CTC, DAMTP. This work was supported by an STFC consolidated grant ST/L000636/1. It was undertaken on the COSMOS Shared Memory system at DAMTP, University of Cambridge operated on behalf of the STFC DiRAC HPC Facility. This equipment is funded by BIS National E-infrastructure capital grant ST/J005673/1 and STFC grants ST/H008586/1, ST/K00333X/1. We acknowledge use of the HEALPix package \cite{Gorski:HEALPix}.
\end{acknowledgments}

\appendix
\section{Magnitude of large eigenvalues of \texorpdfstring{$\mat{\mathscr{C}}$}{normC}}
\label{app:mageigen}
\begin{figure}
\centering
\includegraphics[width=.9\columnwidth]{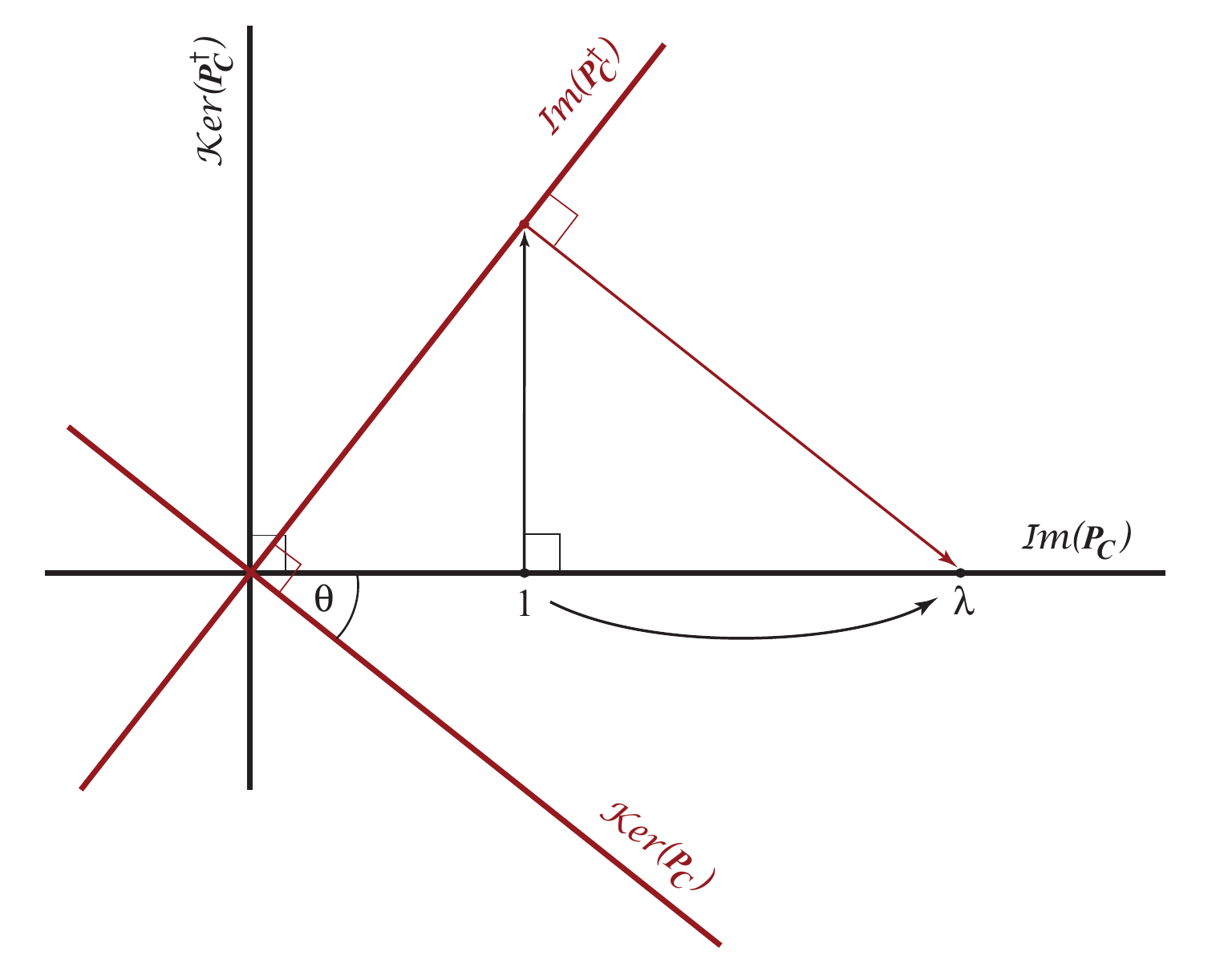}
\caption{Images and kernels of the projection operators $\mat{P}_C$. The successive oblique projections lead to a large eigenvalue $\lambda$.}
\label{fig:2dproj}
\end{figure}

In this appendix we will derive an estimate for the order of magnitude of the large eigenvalues of the normalised covariance $\mat{\mathscr{C}}$ following Ref.~\cite{Gruetjen:TowardsEfficient}. For the purpose of this analysis we slightly alter the definition of the normalised covariance in Eq.~\eqref{eq:normcovdef} to be
\begin{equation}
\mathscr{C}_{l_1m_1l_2m_2}=\frac{P_{l_1m_1l_3m_3}\bar{C}_{l_3}P_{l_3m_3l_2m_2}}{(\bar{C}_{l_1}\bar{C}_{l_2})^{\frac{1}{2}}}\,,
\end{equation}
i.e.\ we simply replace $C_l$ with $\bar{C}_l$ in the numerator. This expression should produce similar large eigenvalues and is identical to the one studied in Ref.~\cite{Gruetjen:TowardsEfficient} giving rise to a simple geometric interpretation with $\bar{C}_l$ instead of $C_l$. We further assume an unapodised mask and include the monopole and dipole in the analysis. To do so, we simply assign values to $\bar{C}_0$ and $\bar{C}_1$, say $\bar{C}_0=\bar{C}_1=\bar{C}_2$. The normalised covariance can then be written exactly in terms of oblique projection operators
\begin{equation}\label{eq:normcovproj}
\mat{\mathscr{C}}=\mat{P}_C\mat{P}_C^{\dagger}
\end{equation}
with
\begin{equation}
(P_C)_{l_1m_1l_2m_2}:=\frac{1}{\bar{C}^{\frac{1}{2}}_{l_1}}P_{l_1m_1l_2m_2}\bar{C}^{\frac{1}{2}}_{l_2}\,,
\end{equation}
where a superscript $^{\dagger}$ denotes the conjugate transpose. For the images and kernels of these projection operators we have
\begin{align}
\mathrm{Im}(\mat{P}_C)&\perp\mathrm{Ker}(\mat{P}_C^{\dagger})\,,\\
\mathrm{Im}(\mat{P}_C^{\dagger})&\perp\mathrm{Ker}(\mat{P}_C)\,.
\end{align}
Figure \ref{fig:2dproj} depicts the emerging geometric picture.
The normalised covariance can develop a very large eigenvector $\lambda$ when $\mathrm{Im}(\mat{P}_C)$ and $\mathrm{Ker}(\mat{P}_C)$ become aligned. Calling the angle between the two $\theta$ it is straightforward to show that
\begin{equation}
\lambda=\frac{1}{\sin^2{\theta}}\,.
\end{equation}
We claimed in Sec.~\ref{subsec:bispecapprox} that a large eigenvector associated with the alignment of image and kernel of $\mat{P}_C$ for a strongly decaying power spectrum is approximately given by
\begin{equation}
(\vec{v}_0)_{lm}=\frac{1}{n_v}\frac{u_{lm}}{\bar{C}_l^{\frac{1}{2}}}\,.
\end{equation}
This vector is obviously an element of $\mathrm{Im}(\mat{P}_C)$ as any eigenvector of $\mat{\mathscr{C}}$ must be according to Eq.~\eqref{eq:normcovproj}. A corresponding element from $\mathrm{Ker}(\mat{P}_C)$ that is very aligned with it is easily found. It is simply given by
\begin{equation}
(\vec{k}_0)_{lm}=\frac{1}{n_k}\frac{u_{lm}-\delta_{l0}\delta_{m0}}{\bar{C}_l^{\frac{1}{2}}}\,,
\end{equation}
where $n_k$ is a suitably chosen normalisation. The angle between $\vec{v}_0$ and $\vec{k}_0$ serves as an upper bound on the minimum angle between $\mathrm{Im}(\mat{P}_C)$ and $\mathrm{Ker}(\mat{P}_C)$. Taking
\begin{equation}
\theta=\arccos{(\mathrm{Re}\lbrace\vec{v}^*_0\cdot\vec{k}_0\rbrace)}\,,
\end{equation}
we obtain an estimate for the large eigenvalue $\lambda$
\begin{equation}
\lambda^3\sim\frac{1}{\sin^6{\theta}}=\left(\frac{1}{1-\mathrm{Re}\lbrace\vec{v}^*_0\cdot\vec{k}_0\rbrace^2}\right)^3\approx 4\times 10^{13}\,.
\end{equation}
When the mask is apodised an interpretation in terms of oblique projections is not exact anymore. However, if we assume that the large eigenvalues can be estimated in the same way by studying the alignment of the vectors $\vec{v}_0$ and $\vec{k}_0$, we can still deduce rough estimates. The results are then obtained by simply replacing the unapodised $u_{lm}$ with the apodised versions. We arrive at estimates $\lambda_{\mathrm{S015}}^3=\mathcal{O}(10^{12})$ and $\lambda_{\mathrm{S05}}^3=\mathcal{O}(10^{10})$.

\section{Apodisation method}
\label{app:ApoMethod}
The apodisation scheme we use in this thesis was designed with several requirements in mind. While the transition of the mask function $U(\hat{\vec{n}})$ from zero in masked regions to unity should be made as smooth as possible to minimise harmonic ringing, we simultaneously want to ensure that regions where the unapodised mask vanishes are also exactly zero in the apodised case so that all masks truly mask the same regions. Furthermore, we aim to minimise the loss of sky fraction as PCL estimators generally lose accuracy with decreasing sky fraction. In the Planck analysis \cite{Ade:2013Likelihood}, the mask is first smoothed with a Gaussian beam, then $0.15$ is subtracted everywhere. All negative values are subsequently set to zero and the resulting mask is scaled by $1/(1-0.15)$ to ensure that it rises to unity in unmasked regions. While this method produces smoothed masks, it is impossible to ensure that the apodised mask vanishes exactly over the same region of the sky as the unapodised mask. As we want to compare different estimators that use the exact same set of data this procedure is not appropriate.

The procedure employed here is based on the following observation. If we convolve a semi-infinite step in the $x$--$y$ plane, i.e.\ the function $f(x,y)=\Theta(x)$, with a Gaussian beam
\begin{equation}
G(x,y)=\frac{1}{2\pi\sigma^2}\exp{\left(-\frac{r^2}{2\sigma^2}\right)}\,,
\end{equation}
where $\sigma=\mathrm{FWHM}/(2\sqrt{2\log{2}})$, the resulting smoothed $f^{\mathrm{smo}}(x,y)$ is given by
\begin{equation}
f^{\mathrm{smo}}(x,y)\equiv S\left(\frac{x}{\mathrm{FWHM}}\right):=\frac{1}{2}\mathrm{erfc}{\left[-2\sqrt{\log{2}}\left(\frac{x}{\mathrm{FWHM}}\right)\right]}\,.
\end{equation}
Performing the remaining steps of the Planck apodisation procedure, we obtain the final apodised mask, $f^{\mathrm{apo}}(x,y)$, as
\begin{equation}
f^{\mathrm{apo}}(x,y)=\begin{cases}\frac{S\left(\frac{x}{\mathrm{FWHM}}\right)-0.15}{1-0.15} & \mbox{for } \frac{x}{\mathrm{FWHM}}>c\ \\ 0 , & \mbox{otherwise} \end{cases}\,.
\end{equation}
Here, $c\approx -0.44013$ is the solution to the equation $\mathrm{erfc}{\left[-2\sqrt{\log{2}}\,c\right]}/2=0.15$. This suggest the following procedure. Rather than actually convolving the binary mask with a Gaussian beam, we determine the distance $r_p$ of a given pixel $\hat{\vec{p}}$ to the closest masked pixel and then assign it the value 
\begin{equation}
U(\hat{\vec{p}})=\frac{S\left(\frac{r_p}{\mathrm{FWHM}}-c\right)-0.15}{1-0.15}\,.
\end{equation}
This procedure gives the same profile as the Planck method if the boundary of the mask is sufficiently straight, while ensuring that no originally masked pixel acquires non-zero weight. Furthermore, we can easily produce masks with different degrees of smoothing.

\bibliography{PhDthesis.bib}

\end{document}